# Systematic Reanalysis of KMTNet microlensing events, Paper I: Updates of the Photometry Pipeline and a New Planet Candidate

Hongjing Yang (杨弘靖)[1]★ , Jennifer C. Yee[2] , Kyu-Ha Hwang[3], Qiyue Qian (钱奇玥)[1] , Ian A. Bond[4], Andrew Gould[5,6], Zhecheng Hu (胡哲程)[1] , Jiyuan Zhang (张纪元)[1] , Shude Mao (毛淑德)[1,7] , and Wei Zhu (祝伟)[1]
(Leading Authors)

Michael D. Albrow[8], Sun-Ju Chung[3], Cheongho Han[10], Youn Kil Jung[3], Yoon-Hyun Ryu[3], In-Gu Shin[2], Yossi Shvartzvald[11], Sang-Mok Cha[3,12], Dong-Jin Kim[3], Hyoun-Woo Kim[3], Seung-Lee Kim[3,9], Chung-Uk Lee[3], Dong-Joo Lee[3], Yongseok Lee[3,12], Byeong-Gon Park[3,9], Richard W. Pogge[3,5], and Weicheng Zang (臧伟呈)[2,1]
(The KMTNet Collaboration)

Fumio Abe[13], Richard Barry[14], David P. Bennett[14,15], Aparna Bhattacharya[14,15], Martin Donachie[16], Hirosane Fujii[17], Akihiko Fukui[18,19], Yuki Hirao[17], Yoshitaka Itow[17,13], Rintaro Kirikawa[17], Iona Kondo[17], Naoki Koshimoto[20,21], Man Cheung Alex Li[16], Yutaka Matsubara[13], Yasushi Muraki[13], Shota Miyazaki[17], Greg Olmschenk[14], Clément Ranc[26] , Nicholas J. Rattenbury[16], Yuki Satoh[17], Hikaru Shoji[17], Stela Ishitani Silva[22,14], Takahiro Sumi[17], Daisuke Suzuki[23], Yuzuru Tanaka[17], Paul J. Tristram[24], Tsubasa Yamawaki[17], and Atsunori Yonehara[25]
(The MOA Collaboration)

[1]*Department of Astronomy, Tsinghua University, Beijing 100084, China*
[2]*Center for Astrophysics | Harvard & Smithsonian 60 Garden St., Cambridge, MA 02138, USA*
[3]*Korea Astronomy and Space Science Institute, Daejon 34055, Republic of Korea*
[4]*Institute of Natural and Mathematical Sciences, Massey University, Auckland 0745, New Zealand*
[5]*Max-Planck-Institute for Astronomy, Königstuhl 17, 69117 Heidelberg, Germany*
[6]*Department of Astronomy, Ohio State University, 140 W. 18th Ave., Columbus, OH 43210, USA*
[7]*National Astronomical Observatories, Chinese Academy of Sciences, Beijing 100101, China*
[8]*University of Canterbury, Department of Physics and Astronomy, Private Bag 4800, Christchurch 8020, New Zealand*
[9]*University of Science and Technology, Korea, (UST), 217 Gajeong-ro Yuseong-gu, Daejeon 34113, Republic of Korea*
[10]*Department of Physics, Chungbuk National University, Cheongju 28644, Republic of Korea*
[11]*Department of Particle Physics and Astrophysics, Weizmann Institute of Science, Rehovot 76100, Israel*
[12]*School of Space Research, Kyung Hee University, Yongin, Kyeonggi 17104, Republic of Korea*
[13]*Institute for Space-Earth Environmental Research, Nagoya University, Nagoya 464-8601, Japan*
[14]*Code 667, NASA Goddard Space Flight Center, Greenbelt, MD 20771, USA*
[15]*Department of Astronomy, University of Maryland, College Park, MD 20742, USA*
[16]*Department of Physics, University of Auckland, Private Bag 92019, Auckland, New Zealand*
[17]*Department of Earth and Space Science, Graduate School of Science, Osaka University, Toyonaka, Osaka 560-0043, Japan*
[18]*Department of Earth and Planetary Science, Graduate School of Science, The University of Tokyo, 7-3-1 Hongo, Bunkyo-ku, Tokyo 113-0033, Japan*
[19]*Instituto de Astrofísica de Canarias, Vía Láctea s/n, E-38205 La Laguna, Tenerife, Spain*
[20]*Department of Astronomy, Graduate School of Science, The University of Tokyo, 7-3-1 Hongo, Bunkyo-ku, Tokyo 113-0033, Japan*
[21]*National Astronomical Observatory of Japan, 2-21-1 Osawa, Mitaka, Tokyo 181-8588, Japan*
[22]*Department of Physics, The Catholic University of America, Washington, DC 20064, USA*
[23]*Institute of Space and Astronautical Science, Japan Aerospace Exploration Agency, 3-1-1 Yoshinodai, Chuo, Sagamihara, Kanagawa, 252-5210, Japan*
[24]*University of Canterbury Mt. John Observatory, P.O. Box 56, Lake Tekapo 8770, New Zealand*
[25]*Department of Physics, Faculty of Science, Kyoto Sangyo University, 603-8555 Kyoto, Japan*
[26]*Sorbonne Université, CNRS, Institut d'Astrophysique de Paris, IAP, F-75014, Paris, France*

## ABSTRACT

In this work, we update and develop algorithms for KMTNet tender-love care (TLC) photometry in order to create an new, mostly automated, TLC pipeline. We then start a project to systematically apply the new TLC pipeline to the historic KMTNet microlensing events, and search for buried planetary signals. We report the discovery of such a planet candidate in the microlensing event MOA-2019-BLG-421/KMT-2019-BLG-2991. The anomalous signal can be explained by either a planet around the lens star or the orbital motion of the source star. For the planetary interpretation, despite many degenerate solutions, the planet is most likely to be a Jovian planet orbiting an M or K dwarf, which is a typical microlensing planet. The discovery proves that the project can indeed increase the sensitivity of historic events and find previously undiscovered signals.

# 1 INTRODUCTION

Gravitational microlensing has been proven to be a powerful tool to detect extrasolar planets (Mao & Paczyński 1991; Gould & Loeb 1992). To date, there are more than 190[1] confirmed microlensing planet detections. With the increasing number of detections, several statistical works that focus on the planet-host mass-ratio function have been presented in this field (e.g., Gould et al. 2010; Suzuki et al. 2016, 2018; Poleski et al. 2021). However, the uncertainties of the current statistical works are still large. For example, Suzuki et al. (2016) suggest that the mass-ratio function has a break at $q_{\rm br} = 1.7 \times 10^{-4}$, but later Jung et al. (2019) argues that the break should be located at a smaller mass ratio, $q_{\rm br} \sim 0.55 \times 10^{-4}$. However, since then, many $q < 10^{-4}$ planets have been discovered[2], which may conflict with a "break" in the mass-ratio function.

The large uncertainty in the statistical results mainly derives from the small number of detections. The reason that the planet numbers in these statistical samples are few is, on the one hand, due to the intrinsically low probability of the microlensing planet perturbation and the difficulty of obtaining sufficiently dense light curve coverage to capture and characterize such perturbations, and on the other hand, potentially due to a failure to find planetary signals in the existing microlensing events. The Korea Microlensing Telescope Network (KMTNet, Kim et al. 2016) has largely solved the former problem by stationing large field-of-view cameras at three sites around the globe, enabling continuous or near-continuous high-cadence observations of 100 deg[2] toward the Galactic bulge. The latter problem can be greatly reduced by applying a semi-automatic search algorithm to the light curves, such as the AnomalyFinder system (Zang et al. 2021b, 2022) of the KMTNet.

However, current searches of KMTNet events are all based on the preliminary data reduced by a real-time or a post-season pipeline (Kim et al. 2018a,b) using the difference image software pySIS (Albrow et al. 2009). This software was adapted into a pipeline that applied to all KMTNet events to produce real-time, photometry posted online (these data are referred to as "preliminary" or "online pipeline" photometry). While the online pipeline includes some adjustments for KMTNet-specific conditions, the pySIS software was meant to be tuned by-hand for the conditions of each individual field. For example, certain sections of the KMTNet camera may have bad pixel columns or some fields may have bright variable stars that need to be masked. Hence, because the online pipeline reduction is not tuned for the conditions in a specific field, by-hand TLC reductions often result in improved results. Therefore, after finding potential signals, the photometry is re-extracted by-hand using a tender-loving care (TLC) approach. This by-hand TLC approach also uses the pySIS software but introduces fine-tuned parameters, special operations, and visual verification of various steps by human operators. The resulting high-quality data are then used for detailed modeling and publication. Therefore, it is possible for the AnomalyFinder and other searches (e.g., by-eye search) based on the preliminary dataset to miss planetary signals that were too subtle but can be enhanced or revealed by the TLC data.

Therefore, we start a project to systematically re-reduce the pho-tometry for the KMTNet microlensing events. In principle, all KMTNet events could be re-reduced using the by-hand TLC procedure, but in practice, these reductions are operator-dependent and time-intensive. The time cost in particular has made it prohibitively expensive to apply to large numbers of KMTNet events, and so precluded a systematic search based on TLC data and also presented challenges in the new era of systematic candidate searches using the AnomalyFinder system (Zang et al. 2021b, 2022).

So, in this project, we start by developing and updating algorithms for KMTNet TLC photometry to enable mostly-automated, high-quality reductions that reproduce the methods of the most highly-skilled operators. These updates form a new TLC pipeline. We plan to apply the new pipeline to the archival events to find possible new planetary signals. Such a systematic reanalysis could both increase the number of planet detections and the survey sensitivity of KMTNet, which will eventually allow us to obtain a more complete and accurate statistical result.

In the first of this series of papers, we describe the development of the new KMTNet TLC pySIS photometry pipeline. Among the challenges for TLC reductions, because the original pySIS algorithm was designed to work on datasets with tens to hundreds of observations for a few events per year, new challenges were encountered once it began to be applied to the hundreds of KMTNet events requiring re-reduction, each with thousands of images. We report on both the underlying principles of the new and updated algorithms, the specific application to KMTNet, and how the changes allow for increased automation, robustness, and improve the accuracy of the pySIS photometry.

As a demonstration, we first apply it to the KMTNet 2019 season prime-field high-magnification events. We report the discovery of a previously undiscovered candidate planetary signal in event MOA-2019-BLG-421 (KMT-2019-BLG-2991), which is revealed by the KMTNet data reduced by the new TLC pipeline. Although there are unresolved degenerate non-planet interpretations, the discovery proves that the procedure can indeed find previously missed anomalies and increase the survey sensitivity.

# 2 PHOTOMETRY PIPELINE

The new TLC pipeline is developed from pySIS (Albrow et al. 2009), which is based on the difference image analysis method (Tomaney & Crotts 1996; Alard & Lupton 1998a; Bramich 2008). The input images are the calibrated images. The process of reducing the photometry with pySIS mainly consists of seven steps:

(a) Preprocess and select an astrometric reference image.
(b) Align images.
(c) Select subtraction reference images and create a master reference image.
(d) Subtract target images from master reference image.
(e) Refine the microlensing source position.
(f) Estimate the Point Spread Function (PSF).
(g) Extract the flux from subtracted images.

We briefly describe the procedures as follows. The purpose of (a) preprocessing is to evaluate the quality of each image, i.e., the full width at half-maximum (FWHM) and the ellipticity of the PSF, the sky background, and the signal-to-noise ratio (SNR). This information is used to define a "good" image. Generally, a "good" image should have small FWHM, small ellipticity, low sky background, and high SNR. One "good" image is chosen to be the astrometric reference image, and all the other images are then aligned to it (step b). For (c), a series of "good" images are stacked to obtain a very high SNR image to use as the master subtraction reference image ($\mathcal{R}$). Then in step (d), a spatially variable numerical kernel $\mathcal{K}$ is solved and used to convolve the reference image to each target image $\mathcal{T}$. The target images are subtracted from the convolved images to obtain the difference images $\mathcal{D}$. In step (e), variable sources show up in the difference images, and therefore the difference images are used to refine the position of the microlensing source (for more details about the algorithm, see Appendix A in Albrow et al. 2009). Then in step (f), the same kernels computed in the subtractions are used to convolve the reference PSF to the target PSF on the final position, to obtain the pixelized PSF models for each target image. Finally, in step (g), a linear fit between each PSF model $\mathcal{P}$ and each difference image $\mathcal{D}$ is performed at the refined source position to extract the flux. Residual images $\mathcal{E}$'s are produced after extracting the flux.

---

[1] https://exoplanetarchive.ipac.caltech.edu, as of 2023-07-05.
[2] For such planets in 2016-2019 events, see Table 14 in Zang et al. (2023). In addition, there are six other planets in 2020-2022 seasons, KMT-2020-BLG-0414Lb (Zang et al. 2021a), KMT-2021-BLG-0171Lb (Yang et al. 2022), KMT-2021-BLG-0912 (Han et al. 2022), KMT-2021-BLG-1391 (Ryu et al. 2022), MOA-2022-BLG-249 (Han et al. 2023a), and KMT-2022-BLG-0440Lb (Zhang et al. 2023).

The by-hand TLC pySIS data reduction procedure of KMTNet was not well optimized for efficiency. Therefore, we carefully reviewed the procedures and made some updates. Many of the challenges in achieving high photometric accuracy come from (c), (e), and (g). In particular, (c) and (e) required significant human input, making it inefficient and the quality of the resulting photometry highly user-dependent; i.e., only certain highly-skilled human operators were able to identify the reasons for bad photometry and correct them. As will be discussed in Sections 2.3 and 2.4, this step can be automated with improved metrics that are optimized for KMTNet data. We note that it is very difficult to make the pipeline both high-quality and fully automated. However, with experience, some common issues can be identified or even prevented by adjusting the algorithm.

The overall procedures have not changed in the new pipeline. However, many detailed operations have been updated.

### 2.1 New Image Quality Metrics

Before we start describing the details of all the updates, we introduce two metrics to evaluate the quality of the photometry. The first is $\sigma_{\rm sub}$, which describes the standard deviation of the subtracted image. For each difference image $\mathcal{D}$, $\sigma_{\rm sub}$ is defined as

$$\sigma_{\rm sub} = {\rm STD}\left(\frac{\mathcal{D}_i}{\sqrt{\mathcal{T}_i}}\right), \quad i \in R_{\rm valid}, \tag{1}$$

where $\mathcal{D}_i$ and $\mathcal{T}_i$ are the $i$-th pixel value of the difference image and the original target image, respectively. $\sqrt{\mathcal{T}_i}$ corresponds to the Poisson noise of the target image. $R_{\rm valid}$ is the region of all unmasked pixels. Here "STD($x_i$)" means the standard deviation of all $x_i$. The quality of the subtraction is described by $\sigma_{\rm sub}$. If an image is well-subtracted, the subtraction residuals $\mathcal{D}_i/\sqrt{\mathcal{T}_i}$ should approximately follow the standard normal distribution, therefore $\sigma_{\rm sub} \sim 1$; a larger value represents a worse subtraction.

The second metric is $\sigma_{\rm res}$, defined as

$$\sigma_{\rm res} = \log_{10}\left(\frac{1}{N_{\rm phot}} \sum_i |\mathcal{E}_i\,\mathcal{P}_i|\right), \quad i \in R_{\rm phot}, \tag{2}$$

where $\mathcal{E}_i$ is the $i$-th pixel value of the residual image, and $\mathcal{P}_i$ is the value of the normalized PSF model on the corresponding pixel. The photometry region is $R_{\rm phot}$ and the number of the pixels in $R_{\rm phot}$ is $N_{\rm phot}$. The quality of the photometry in $R_{\rm phot}$ is described by $\sigma_{\rm res}$; larger $\sigma_{\rm res}$ values represent worse photometry. Because stars have different brightnesses, backgrounds, and blend levels, the expected $\sigma_{\rm res}$ value varies, but generally, $\sigma_{\rm res} \gtrsim 2$ indicates poor photometry.

These two metrics help us to quantify the goodness of image subtraction and flux extraction. All the updates below are based on these two measures, that is, each update makes at least one of $\sigma_{\rm sub}$ or $\sigma_{\rm res}$ smaller. The two metrics also help us to identify problematic data points caused by poor original images or errors in the photometry process.

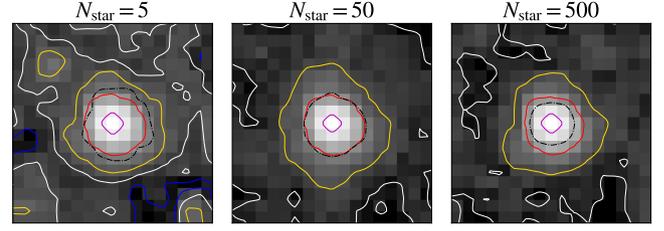

**Figure 1.** PSF images of an example image, extracted using number of combined stars $N_{\rm star} = (5, 50, 500)$. The PSFs are normalized and presented in log scale. The (blue, white, yellow, magenta) lines represent the $(-0.001, 0, 0.001, 0.01, 0.1)$ contours. The dashed black line is the contour of SNR = 5. When $N_{\rm star}$ is too small (left), the background pixels are noisy because of small number statistics. When $N_{\rm star}$ is too large (right), the SNR of the center pixels are lower because they are dominated by faint stars. We find for $2' \times 2'$ image stamps of the KMTNet bulge field, $N_{\rm star} = 30 \sim 200$ is the best (middle).

### 2.2 PSF Extraction

Because it affects multiple steps, we start by describing changes to the algorithm to extract the PSF. The PSF extraction is important in two aspects. First, accurate PSF estimation can help us identify the quality of each original image, thus can help to find better reference image(s). Second, the PSF of the reference image is used to create the PSF model of all target images. Poor PSF estimation will cause poor flux extractions on all images. Therefore, we updated the PSF extraction script to compute the PSF more accurately .

The original pySIS used the `Bphot` script in `ISIS`[3] (Alard & Lupton 1998b; Alard 2000) to compute a pixelated PSF for a given image. First, the script detects several ($N_{\rm star}$) of the brightest stars that are not saturated. Then, it median combines them to obtain a clean PSF that eliminates crowded neighboring stars. When $N_{\rm star}$ is too small, the background pixels are noisy because of small number statistics. When $N_{\rm star}$ is too large, the SNR of the center pixels are lower because they are dominated by faint stars. We find for $2' \times 2'$ image stamps of the KMTNet bulge field (the PSF does not change much on this scale), $N_{\rm star} = 50$ is a better default choice than previous default number $N_{\rm star} = 5$. In the rare cases when 50 is not optimal, the value is usually in the range $N_{\rm star} = 30 \sim 200$. See Fig. 1 for an example.

### 2.3 Preprocessing of Images

The primary purpose of the preprocessing is to evaluate the quality of all original images and provide information for reference image selections. The original pySIS uses a simple algorithm to estimate the FWHM, ellipticity, and the sky background. In the case of FWHM and ellipticity, the PSF is estimated by linearly fitting the two-point correlation function to a Gaussian function along the image axis X and Y, respectively, denoted as FWHMX and FWHMY.

---

[3] http://www2.iap.fr/users/alard/package.html

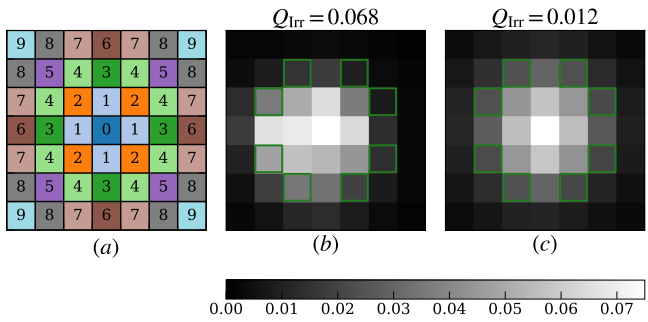

**Figure 2.** ($a$). Schematic of the $Q_{\text{Irr}}$ definition. The numbers in each pixel represents the index $k$, where pixels with identical $k$'s have the same distance to the central $k = 0$ pixel. The colors are arbitrary, but indicate pixels with the same $k$. The final irregularity, as defined in Eq. 3, is $Q_{\text{Irr}} = \text{STD}(\mathcal{P}_{k=0}) + \text{STD}(\mathcal{P}_{k=1}) + \text{STD}(\mathcal{P}_{k=2}) + \cdots$. ($b$) and ($c$). Examples of irregular and symmetric PSF images, respectively. To illustrate, $k = 4$ pixels are highlighted with green boxes. The values of $k = 4$ pixels are basically identical in ($c$) but are significantly different in ($b$). The final $Q_{\text{Irr}}$ values are labeled on the top of each panel.

This algorithm is fast and efficient, but also has a number of limitations. For example, the ellipticity is estimated by FWHMX/FWHMY. The algorithm can incorrectly estimate the ellipticity and FWHM, particularly when the ellipse's axis ratio is large and its major axis is angled at close to $\pm 45°$. In addition, the simple algorithm uses all the pixels on the image and generally overestimates the FWHM by about $\sim 1$ pixel, resulting from saturated stars and bleeding spikes that are very common in KMTNet images. Moreover, the sky background is simply estimated by the 85-th percentile of all pixel values.

Therefore, we introduce DoPhot (Schechter et al. 1993) to accurately estimate the FWHM, ellipticity, and the sky background.

We also add an image quality indicator about the PSF irregularity as a supplement to the ellipticity. We use the modified version of Bphot in Section 2.2 to extract the PSF image, and then measure the irregularity $Q_{\text{Irr}}$ by summing the standard deviations over every "annulus" of the PSF

$$Q_{\text{Irr}} = \sum_{0 \leqslant r_k \leqslant r_{\text{max}}} \left[ \sum_{x_i^2 + y_j^2 = r_k^2} \text{STD}(\mathcal{P}_{i,j}) \right]. \tag{3}$$

Where STD is the standard deviation of the given pixels and $r_{\text{max}}$ is the radius of the PSF image, by default 2.5 FWHM. A schematic can be seen in Fig. 2($a$). If the PSF is completely symmetric, $Q_{\text{Irr}}$ will be zero because all the points with the same radius will be equal. Thus larger $Q_{\text{Irr}}$ values represent larger irregularity.

This process takes significantly longer to run than the previous algorithm (a few minutes rather than a few seconds). However, as we will see in the next section, these quantities, mainly FWHM, sky background, $Q_{\text{Irr}}$, and the standard deviation of each image, enable us to automatically select high-quality reference images without human decisions.

## 2.4 Selection of Reference Images

In the original by-hand TLC procedure, a human reviewer was required to check the quality of reference images. This was especially important for a dataset the size of KMTNet, because relatively rare edge cases (false "good" images) dominated the "best" images se-lected using the metrics calculated according to the original method. The result was that reference images often had to be selected manually with only limited information to assess why one image (or set of images) was better than another. This sometimes required repeating the entire process multiple times with different combinations of reference images to test if there was any improvement in the resulting photometry. With the more accurate and expanded metrics described in the previous section, we can both improve and better automate the reference image selection.

The image alignment and the overall photometry are not sensitive to the selection of the astrometric reference image. Therefore, here we only focus on the subtraction reference image(s) selection. Hereafter, unless specified, "reference images" refers to the images that are used to stack into the final, single, "master reference image" $\mathcal{R}$ used in the subtraction.

Because the master reference image is used to convolve and subtract all the other images, it requires the reference image(s) to have

- high SNR. Otherwise the convolution kernels cannot be accurately determined, which would affect the reliability of the subsequent subtractions and flux measurements.
- small FWHM. Because a larger FWHM image is harder to convolve to a smaller FWHM one. The reference images should be roughly the best seeing images.
- symmetric PSF. Similar to the above requirement, an asymmetric PSF cannot be convolved to symmetric PSFs with similar FWHM.

In addition, we want the target source to have approximately the same flux in the set of reference images because stacking variable sources can introduce extra systematic errors.

Therefore, based on the information we obtain from Section 2.3, we select the smallest FWHM, highest SNR, and smallest $Q_{\text{Irr}}$ images. In detail, we start by setting thresholds on SNR > 15%, sky background < 80%, $Q_{\text{Irr}} < 0.03$, and $T \notin [t_0 - 20 \text{ d}, t_0 + 20 \text{ d}]$, where $T$ is the observation time of each image, and $t_0$ is the peak time of the microlensing event. The thresholds, especially the time interval, can be changed for different events. Here we only present the typical numbers. The thresholds typically rule out $\sim 40\%$ of images. The remaining images are then sorted by FWHM, and the best $\sim 40$ are selected as the initial reference images.

Then, among these initial reference images, we start an iterative process. First, the 20 best FWHM images are convolved to the initial reference image and then averaged into an initial master reference image, which is then subtracted from the best 40 images. After the subtraction, the 20 images with the best $\sigma_{\text{sub}}$'s are selected as the reference images for the next iteration. Usually, after 3-5 iterations, the process converges to 20 images, which are taken as the final reference images. They are stacked into the final master reference image, $\mathcal{R}$. Note that after this process, the master reference image is not necessarily constructed from the best seeing or lowest ellipticity images, but the algorithm is robust at rejecting bad images from being used to create the master reference image. With this new reference image selection algorithm, it becomes possible to automate this step without human interaction.

## 2.5 Image Subtraction

The image subtraction algorithm follows Albrow et al. (2009) and is not updated. We only modify the script to allow for adding extra masks and account for images with FWHM smaller than the master reference image PSF.

The kernel can only describe the difference between the reference PSF and target PSF if all of the stars on the images are constant.

Therefore, we need to mask the non-constant stars when calculating the kernel. (After the kernel is solved, all pixels are convolved and subtracted.)

However, the default pySIS mask only contains saturated pixels and zero-value pixels. It does not detect and mask variable stars or bad pixels. Therefore bad pixels and variable stars can produce deviations in the kernel $\mathcal{K}$ and consequently on the difference image $\mathcal{D}$. See Figure 3 for examples. Therefore, we have updated the script to allow it to use a global mask for all images and individual masks for each image.

Global masks are usually used for masking variable stars. The fluxes of these stars vary considerably from image to image. If they are not masked, the kernel and thus the subtraction will be unreliable. Because the variable stars are at the same pixel coordinates after the image alignments, a global mask for all images is able to handle it. On the other hand, individual masks are usually used for CCD bad columns and other CCD defects. Those pixels are not at fixed coordinates on the aligned images but have the same position on the CCD, therefore they need to be individually masked in all images.

The variable stars can be automatically detected during the iteration in Section 2.4 by averaging over the absolute value of the difference images in the reference set. For constant stars, the averages are dominated by the Poisson error and thus small, but for variable stars the variable fluxes from different phases will add up. After the averaging, any pixels above 1000 counts[4] are masked. Although it is possible that some variable stars occasionally have similar flux over the reference set, this procedure can find most of them automatically. Global masks of these variable stars are then applied to all the subsequent subtractions.

In addition to the masks, although we have optimized the reference image selection, unavoidably there remain some images that cannot be well subtracted. They are the images with smaller FWHM (along both axes or only the minor axis of the PSF) than the reference image. For these images, we first convolve (blur) them to a intermediate image $\mathcal{T}'$ by a normalized Gaussian kernel $\mathcal{K}_\sigma$. Therefore, the optimization of the kernel $\mathcal{K}$ becomes

$$\mathcal{R} \otimes \mathcal{K} \rightarrow \mathcal{T}' = \mathcal{T} \otimes \mathcal{K}_\sigma. \quad (4)$$

The Gaussian kernel size $\sigma$ is determined by

$$(2\sqrt{2\ln 2}\sigma)^2 + \mathrm{FWHM}_{\mathrm{minor}}^2 = \mathrm{FWHM}_{\mathrm{ref,major}}^2, \quad (5)$$

where $\mathrm{FWHM}_{\mathrm{minor}}$ is the minor-axis FWHM of the original target image $\mathcal{T}$, and $\mathrm{FWHM}_{\mathrm{ref,major}}$ is the major-axis FWHM of the reference image $\mathcal{R}$. The constant $2\sqrt{2\ln 2} \approx 2.355$ comes from the FWHM of a standard Gaussian function.

### 2.6 Source Position Refinement

The accuracy of the source position can directly affect the quality of the photometry (for example, see Fig. 1 in Albrow et al. 2009). We use the algorithm by Albrow et al. (2009) to calculate the source position and its error for all individual (subtracted) images, but we update the algorithm for determining the source position from multiple measurements.

We denote the measured source position of the $k$-th image as

$\bar{x}_k = (x_k, y_k)$, and its error as $(\sigma_{x,k}, \sigma_{y,k})$. The weight of the $k$-th measurement is then

$$w_k = \frac{1}{\sigma_{x,k}^2 + \sigma_{y,k}^2}. \quad (6)$$

We first exclude $\sigma_{\mathrm{sub}} > 2.0$ images. These badly subtracted images sometimes provide nominally precise but incorrect position measurements. For the remaining images, we start an iteration. In each iteration, we compute the weighted centroid $\bar{x}_c = (x_c, y_c)$, where

$$x_c = \sum_k w_k x_k \Big/ \sum_k w_k, \quad (7)$$

$$y_c = \sum_k w_k y_k \Big/ \sum_k w_k, \quad (8)$$

and its covariance $\mathcal{C}$,

$$C = \begin{pmatrix} \frac{\sum_k w_k (x_k - x_c)^2}{\sum_k w_k} & \frac{\sum_k w_k (x_k - x_c)(y_k - y_c)}{\sum_k w_k} \\ \frac{\sum_k w_k (x_k - x_c)(y_k - y_c)}{\sum_k w_k} & \frac{\sum_k w_k (y_k - y_c)^2}{\sum_k w_k} \end{pmatrix}. \quad (9)$$

Then all $> 3\sigma$ points

$$\sqrt{(\bar{x}_k - \bar{x}_c) C^{-1} (\bar{x}_k - \bar{x}_c)} > 3 \quad (10)$$

are excluded. The remaining images are used in the next iteration. The iteration continues until all remaining images are within $3\sigma$.

### 2.7 Photometry Flux Extraction

In pySIS, the reference PSF is convolved to each target image and interpolated to the refined source position. For each original image, we now have the subtracted difference image $\mathcal{D}$ and the normalized pixelated PSF model $\mathcal{P}$. The flux $f$ is then the slope of the linear fit between $\mathcal{D}$ and $\mathcal{P}$.

For KMTNet photometry, we introduce an additional parameter, b, such that

$$\mathcal{D} = f \cdot \mathcal{P} + b, \quad (11)$$

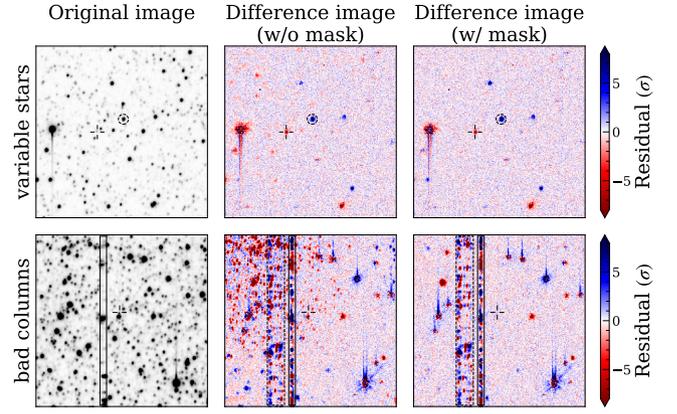

**Figure 3.** Example images showing the influence of variable stars (upper panels) and bad columns (lower panels) with and without masking. The left, middle, and right column shows the original images, difference images without masking, and difference images with the extra mask, respectively. The difference images show the subtraction residuals (divided by the Poisson noise). In all panels, the cross marks the location of the microlensing event. The dashed circle on the top panels indicates the variable star. The region between vertical black lines on the lower panels are the location of the CCD bad columns, and the dashed regions are the bad columns on the reference image.

---

[4] The value 1000 is a default value and is valid in most cases in practice. It can be changed manually.

where $b$ is a free parameter describes the "background" of the difference image. In most cases, the background of $\mathcal{D}$ is close to zero, and $b$ is consistent with zero. In a few cases, mainly for large FWHM difference images, the backgrounds or local backgrounds around the source significantly deviate from zero, and $b$ can reproduce the deviations.

In addition, pySIS does not use all pixels in $\mathcal{D}$ and $\mathcal{P}$ in the linear fit. Pixels far from the center do not contain any information about the star but only noise. We only use pixels in a circle centered at the source position. The default pySIS used a fixed radius of 6 pixels for this circle. After tests with KMTNet images, in our new TLC pipeline, we adopt a FWHM-related radius

$$r_{\rm phot} = 1.5 \times {\rm FWHM~(pixels)}, \tag{12}$$

with the minimum and maximum value of $(6, 20)$. We denote this region as $R_{\rm phot}$.

For the linear fit with the addition of the $b$ parameter, we minimize

$$\chi^2 = \sum_i \frac{(\mathcal{D}_i - f \cdot \mathcal{P}_i + b)^2}{\sigma_i^2} \tag{13}$$

where $i$ represents $i$-th pixel in the region $R_{\rm phot}$ and $\sigma_i^2$ is the noise of pixel $i$,

$$\sigma_i^2 = \mathcal{T}_i + \sigma_{\rm RON}^2 + \sigma_{\rm other}^2, \tag{14}$$

where $\mathcal{T}$ is the original target image, $\mathcal{T}_i$ is the corresponding $i$-th pixel value (i.e., the Poisson variance of the pixel value). $\sigma_{\rm RON}^2$ is the read-out noise, and $\sigma_{\rm other}^2$ is the other unrecognized noise initialized at zero. The Poisson noise of the reference image is ignored since it is negligible compared to $\mathcal{T}$.

So, in our case, the result of the linear fit can be written analytically (more on linear fitting, see also Gould 2003),

$$\begin{pmatrix} b \\ f \end{pmatrix} = \begin{pmatrix} \langle 1 \rangle & \langle \mathcal{P} \rangle \\ \langle \mathcal{P} \rangle & \langle \mathcal{P}^2 \rangle \end{pmatrix}^{-1} \begin{pmatrix} \langle \mathcal{D} \rangle \\ \langle \mathcal{P}\mathcal{D} \rangle \end{pmatrix}, \tag{15}$$

where we denote the average of a quantity $A$ as

$$\langle A \rangle = \sum_{i \in R_{\rm phot}} \left( \frac{A_i}{\sigma_i^2} \right). \tag{16}$$

The variances of the parameters are

$$\begin{pmatrix} \sigma_b^2 \\ \sigma_f^2 \end{pmatrix} = {\rm diag} \begin{pmatrix} \langle 1 \rangle & \langle \mathcal{P} \rangle \\ \langle \mathcal{P} \rangle & \langle \mathcal{P}^2 \rangle \end{pmatrix}^{-1}. \tag{17}$$

This is a change from the default pySIS, which uses

$$\sigma_f^2 = (\textstyle\sum_{i \in R_{\rm phot}} \mathcal{P}_i)^2 / \langle \mathcal{P}^2 \rangle. \tag{18}$$

To eliminate the impact of cosmic rays, bad pixels, and other noise sources, pySIS excludes pixels with $|\mathcal{D}_i - f \cdot \mathcal{P}_i + b| / \sigma_i > 2.5$ pixels from the fitting (with a slight modification to include our $b$ parameter).

Considering that there might be unrecognized extra noise sources, we calculate the $\sigma_{\rm other}^2$ term to make the reduced chi-square

$$\chi_{\rm red}^2 = \frac{\chi^2}{N_{\rm phot} - 2} \leqslant 1, \tag{19}$$

where $N_{\rm phot}$ is the number of pixels used in the linear fit thus $N_{\rm phot} - 2$ is the number of degrees of freedom for the linear fit. We require

$\sigma_{\rm other}^2 \geqslant 0$[5]. We then iterate the whole fitting process until $\sigma_{\rm other}^2$ and the valid pixels are converged.

## 2.8 Summary and Application

The updates described in Sections 2.1 − 2.7, fall into four major categories. First, we now compute FWHM, ellipticity, and sky background using DoPhot and we have added additional metrics ($Q_{\rm irr}$, $\sigma_{\rm sub}$, $\sigma_{\rm res}$) to assess the image quality and quality of the image subtraction and photometry. With those metrics, we can automate the reference image selection, and we have also automated the process for masking pixels and identifying the source position. Third, we have found that for KMTNet data, using $N_{\rm star} = 50$ works better for computing the PSF. Finally, we have made a few modifications to the pySIS algorithm. We find that it is better to allow let the background of the difference image, $b$, be a free parameter and to use a radius proportional to the FWHM when extracting the flux. We also convolve small FWHM images to the FWHM of the master reference image and change the calculation of the flux uncertainty.

In summary, most of these modifications are aimed at allowing the photometry extraction to be automated and remove the humandependent factors from the photometry. However, changes in the last category can significantly reduce seeing correlations in the data, which can result in improved photometry over the original pySIS for some subset of datasets. A more detailed discussion of the accuracy and efficiency of the updates can be found in Section 7.

The new TLC pipeline has been applied to more than 100 events[6], including > 50 for the final analysis of known anomalous events, and ~ 50 for the systematic search. Among them, 21 have been published. The published events are listed in Table 1.

In this paper, we report a newly discovered candidate planetary signal in microlensing event MOA-2019-BLG-421. The light curve together with the point-source point-lens (PSPL) Paczyński (1986) model is shown in Figure 4. The left panels show the original online pySIS light curve, and the right panels are the re-reduced light curve utilizing the updated algorithm. From the comparison, it is clear that the data from the new TLC pipeline have significantly lower scatter than the online pipeline data. The light curve becomes smoother with less scatter. Moreover, the new TLC pipeline can detect problematic data points by $\sigma_{\rm sub} > 2.5$ or $\sigma_{\rm res} > 2.0$, which are the gray "x" points in the right panels.

With the new data, we find a subtle asymmetric signal relative to the peak of the light curve. The most obvious anomalous regions $T_1$ and $T_4$ are marked on the lowest panels of Figure 4. In the new light curve (lower right panel), the data in $T_1$ is significantly above the PSPL model and the data in $T_4$ are below the PSPL model. The online pySIS light curve also indicates anomalies over $T_1$ and $T_4$. However, the signals are at the same level or even weaker than other features that are, in fact, due to systematic errors, e.g., $T_2$ and $T_3$, even if the scattered KMTA03 data are removed. This is the reason why the AnomalyFinder algorithms (Zang et al. 2021b, 2022) identified it as a noisy event and failed to find the real signal in the online data. The analysis of this signal is presented in Section 4.

---

[5] Mathematically, it is possible to have $\sigma_{\rm other}^2 < 0$ because we fit for the whole $\sigma_{\rm other}^2$ term rather than $\sigma_{\rm other}$ itself.
[6] As of June 2023.

## 3 OBSERVATION OF MOA-2019-BLG-421

The microlensing event MOA-2019-BLG-421 is located in the Galactic bulge. It was first discovered by the Microlensing Observations in Astrophysics (MOA, Bond et al. 2001; Sumi et al. 2003) collaboration on 2019-09-17. The event was also indentified by the post-season KMTNet EventFinder system (Kim et al. 2018b) and named KMT-2019-BLG-2991. Hereafter, we use the name MOA-2019-BLG-421 following the first discovery. The equatorial and Galactic coordinates of the event are (R.A., Dec.)$_{J2000}$ = (18 : 06 : 10.91, −27 : 29 : 07.69) and $(l, b)$ = 3.534°, −3.184°, respectively. The event was located in MOA-GB14 field and two slightly offset KMTNet fields, BLG03 and BLG43, leading to a combined cadence of $\Gamma \sim 4 hr^{-1}$. This cadence covered the peak of the event well.

The images from the MOA survey were mainly taken in the MOA-red wide band, which is approximately the sum of the standard Cousins $R$ and $I$ bands, and a fraction of images taken in the $V$ band. The majority of images of the KMTNet survey were taken in the $I$ band, and about 9% were taken in the $V$ band for color measurements.

The data used in the light-curve analysis were reduced using various difference image analysis pipelines. The MOA data were reduced by Bond et al. (2001). The KMTNet $I$ band data were first reduced by the original KMT pySIS pipeline for producing preliminary, on-line photometry (Albrow et al. 2009), and then reduced by the new TLC pipeline described in Section 2. In addition, we conduct pyDIA[7] photometry to measure the source color in KMTC03 $I$- and $V$- band images. The pySIS flux is then calibrated to the pyDIA flux.

The anomaly of the event is well characterized by the KMTNet TLC data. It was found by using only the KMTNet data. The MOA data alone could not independently discover the anomaly because of the sparse coverage, however, it supports the discovery from KMT-Net.

## 4 LIGHT-CURVE ANALYSIS OF MOA-2019-BLG-421

The light curve shown in Figure 4 was first fitted by a standard single-lens single-source (1L1S) model. The 1L1S model consists of at least four parameters $(t_0, u_0, t_E, \rho)$, where $t_0$ is the time when the lens and the source are closest, $u_0$ is the impact parameter in the units of the angular Einstein radius $\theta_E$ of the total lens mass, $t_E$ is the Einstein radius crossing time or microlensing timescale, and $\rho$ is the radius of the source star in the units of $\theta_E$. In addition, two flux parameters are needed $(f_{S,i}, f_{B,i})$ for each data set $i$, representing the flux of the source and the blend. The fitting parameters and their uncertainties for the 1L1S model are shown in Table 2. As might be anticipated, there is only an upper limit on $\rho \lesssim u_0$.

From the lower right panel of Figure 4, one can discern a weak residual from the standard 1L1S model. The left wing of the peak (HJD′ $\sim 8741 − 8742$) is slightly brighter than the 1L1S model, and the right wing (HJD′ $\sim 8744 − 8748$) is slightly fainter than the model. The signals are subtle, however, due to the relatively long duration of the anomaly and the good coverage from all three KMT-Net observations, those data points actually contribute $\Delta \chi^2 \sim 180$, which is significant enough for a reliable detection.

Such a signal can potentially be reproduced by many models apart from the standard 1L1S model. Below we separately discuss the possible models, including 1L1S with higher-order effects, binary source (1L2S) models, and binary lens (2L1S) models.

### 4.1 1L1S with Microlensing Parallax

The microlensing parallax effect caused by the orbital motion of Earth (Gould 1992, 2000, 2004) can create asymmetry in the light curve. The microlens parallax is

$$\vec{\pi}_E = \frac{\pi_{rel}}{\theta_E} \frac{\vec{\mu}_{rel}}{\mu_{rel}}, \quad \pi_{rel} = AU \left( \frac{1}{D_L} - \frac{1}{D_S} \right) \quad (20)$$

where $(\pi_{rel}, \vec{\mu}_{rel})$ are the lens-source relative parallax and proper motion, and $(D_L, D_S)$ are the distances from Earth to the lens and the source, respectively.

We add two parallax parameters $\pi_{E,E}$ and $\pi_{E,N}$ (east and north component of $\vec{\pi}_E$) to the model. The ecliptic degeneracy (Smith et al. 2003; Jiang et al. 2004; Skowron et al. 2011) is considered by fitting $(u_0 > 0, u_0 < 0)$ scenarios separately. The results are shown in Table 2. The $\chi^2$ is improved by 73.5 for both $u_0 > 0$ and $u_0 < 0$ cases.

However, the measurement of $\vec{\pi}_E$ for such a $t_E \sim 15d$ short-timescale event is uncommon, thus we investigate the solutions carefully. We find the parallax signal is not self-consistent as a function of time, i.e., the signals before and after the peak are in conflict. Taking $u_0 > 0$ as an example, the pre-peak is better by $\Delta \chi^2 = \chi^2_{parallax} - \chi^2_{static} \sim −105$ but the post-peak is worse by $\Delta \chi^2 \sim 35$. Moreover, from Section 4.4, we note that even the best 1L1S + parallax model is disfavored by $\Delta \chi^2 > 70$ relative to the xallarap solution and all the 2L1S solutions described below. Therefore, we exclude the 1L1S + parallax scenario.

### 4.2 Static 1L2S

A second source that is fainter or has a larger impact parameter could also produce an asymmetric feature in the peak. To model the standard 1L2S light curve, three 1L1S parameters of the second source plus one flux parameter in each band are needed (Hwang et al. 2013). We use $(t_{0,1}, u_{0,1}, \rho_1)$ as the 1L1S parameters of the primary source, and $(t_{0,2}, u_{0,2}, \rho_2)$ as the impact time, impact parameter, and the size of the second source. The two sources share the same timescale $t_E$. For each observational band $i$, we use $q_{F,i}$ as the flux ratio between the second source and the primary source.

The results are shown in Table 2. The model and its residuals are shown in Figure 5. The 1L2S model mainly improves the fitting before the peak, but still left residuals over the peak ($t \sim 8743.1 − 8743.8$) and after the peak ($t \sim 8744.8 − 8748.5$). From Section 4.4, we find the 1L2S model is disfavored by $\Delta \chi^2 \gtrsim 50$ with respect to the 2L1S solutions. Therefore, the 1L2S model cannot describe the data well, and we exclude this scenario. Moreover, we also check for parallax effects. However, for similar reasons as in Section 4.1, the signal is inconsistent over time, and we exclude the 1L2S + parallax scenario as well.

### 4.3 1L1S with Microlensing Xallarap

The microlensing "xallarap" effect is caused by the orbital motion of the source star (Griest & Hu 1992; Han & Gould 1997), which we later see is consistent with the data. Such motion would cause the primary source to be accelerated, and thus could produce an asymmetric peak in the light curve.

We consider a circular orbit for the source. The xallarap model introduces five more parameters (Miyazaki et al. 2021), the period of the orbital motion $P_\xi$, the orbital inclination $i_\xi$, the orbital phase $\phi_\xi$, and the xallarap parameter $\vec{\xi}_E$. The phase $\phi_\xi$ is the phase of the source when $t = t_0$, while $\phi_\xi = 0$ is defined as the case where the source is at the ascending/descending node. The amplitude of the xallarap $\xi_E$ is the semi-major axis of the source star orbit normalized by $\hat{r}_E$, the angular Einstein radius projected to the source plane. The direction of $\vec{\xi}_E$ is defined as $\theta_\xi$, the angle between the linear lens trajectory and the orbital ascending/descending node in the range of $[0, \pi)$.

For events with $u_0 \ll 1$ like MOA-2019-BLG-421, the xallarap model has a pair of degenerate solutions,

$$\begin{cases} i_{\xi,1} + i_{\xi,2} \sim \pi \\ |\phi_{\xi,1} - \phi_{\xi,2}| \sim \pi \\ \xi_{E,E,1} - \xi_{E,E,2} \sim 0 \\ \xi_{E,N,1} + \xi_{E,N,2} \sim 0 \end{cases} \quad (21)$$

where the subscript "1" and "2" denote the two degenerate solutions. This degeneracy arises from the geometric symmetry of the trajectories when $u_0 \to 0$. We start by searching for the first solution and then use Eq. 21 to find the degenerate solution.

We search for the local $\chi^2$ minima in the xallarap period $P_\xi$ space. We select a series of $P_\xi$ values from 4 to 500 days and fix them when optimizing. All the other parameters including the PSPL and xallarap parameters are set free. For each $P_\xi$, we search both degenerate solutions. The search results are shown in Figure 6. The lower panel of Figure 6 shows that the $\chi^2$ as a function of $P_\xi$ is smooth, and the only $< 3\sigma$ local minima is $P_\xi = 12.6$ d. This results in only one pair of solutions.

Table 3 shows the final optimized parameters of the xallarap models. We denote the two degenerate solutions "+" and "-" by their $\xi_{E,N}$ sign. The xallarap models describe the light curve significantly better than the static 1L1S, 1L1S + parallax, and static 1L2S models by $\Delta\chi^2 \sim (193, 119, 88)$. The light curve and model residuals can be found in Figure 5. However, we notice that $\Delta\chi^2 \sim 20$ is contributed by the data taken around the full-moon nights (HJD' $\sim 8734 - 8737$, i.e., $\tau = -0.75 \sim -0.5$), which could be caused by systematic errors. Therefore, we also report the $\chi^2_{\text{peak}}$ for the high signal-to-noise $t_0 \pm 4.5$ d peak region in Table 3. This will be used to compare with the 2L1S model below.

Now, we simply check whether the system is physically reasonable. Because the parameters related to the physical properties ($P_\xi, \xi_E$) are consistent within $1\sigma$ for the two degenerate solutions, we take the "−" solution as an example. The results show that the source is in a binary system with a period of $\sim 14.3$ d. The xallarap amplitude is $\xi_E = \sqrt{\xi_{E,N}^2 + \xi_{E,E}^2} = 0.061 \pm 0.009$. The semi-major axis of the source is

$$a_S \equiv \xi_E D_S \theta_E = 0.136 \xi_E \text{ AU} \left( \frac{D_S}{8.3 \text{ kpc}} \right) \left( \frac{\mu_{\text{rel}}}{6 \text{ mas/yr}} \right) \left( \frac{t_E}{\text{d}} \right). \quad (22)$$

Assuming that the mass ratio between the companion and the source is $q_S$, we can relate the total mass of the system to the observables using the Kepler's third law,

$$M_{\text{tot}} \equiv M_S(1 + q_S) = \left( \frac{a_{\text{tot}}}{\text{AU}} \right)^3 \left( \frac{P_\xi}{\text{yr}} \right)^{-2} M_\odot \quad (23)$$

where $a_{\text{tot}} = a_S(1 + q_S)/q_S$. We denote a dimensionless parameter $Z$ by combining Eqs. 22 and 23, where

$$Z \equiv \left( \frac{a_S}{\text{AU}} \right)^3 \left( \frac{P_\xi}{\text{yr}} \right)^{-2} \left( \frac{M_S}{M_\odot} \right)^{-1} = \frac{q_S^3}{(1 + q_S)^2}. \quad (24)$$

For typical microlensing sources $D_S = 8.3$ kpc and $M_S = 1 M_\odot$, we plug the values and the model parameters of the "−" solution into Eq. 24,

$$Z \approx 0.60 \left( \frac{\mu_{\text{rel}}}{6 \text{ mas/yr}} \right)^3. \quad (25)$$

By assuming the mass ratio $q_S = 0.5$, we estimate $\mu_{\text{rel}} = 2.7$ mas/yr, which is a common value for the Galactic bulge stars. Therefore, the system is reasonable given Galactic dynamics.

In addition, we do not consider the flux contributed by the companion in the modelling. We now check whether this assumption is self-consistent. The model and the assumed mass ratio gives $a_{\text{tot}}/D_S = 0.18\theta_E$, assuming the source and the companion are both normal main-sequence stars, the companion should be both fainter ($f_C/f_S \sim q_S^4 \sim 0.06$) and farther from the magnification center ($a_{\text{tot}}/D_S/\theta_E \sim 7 u_0 \gg u_0$). As a result, the companion contributes $< 1\%$ magnified flux to the peak. Thus, neglecting the companion flux is self-consistent.

Therefore, we conclude that the xallarap solutions are physically reasonable. We keep these models as one of the final interpretations.

### 4.4 2L1S

Binary-lens microlensing (Mao & Paczyński 1991; Gould & Loeb 1992), because of its non-linearity, can produce diverse light curves including asymmetric ones. In this event, the anomaly is centered at the peak. This feature indicates that it can be caused by the perturbation of the central caustic or cusp (e.g., Chung et al. 2005).

We use three extra parameters in addition to the 1L1S scenario to model the 2L1S light curves. They are $s, q, \alpha$, where $s$ is the projected distance between the two lenses in the units of $\theta_E$, $q$ is the mass ratio of the two lenses, and $\alpha$ is the angle between the source trajectory and the binary lens axis in the lens plane.

Because the 2L1S parameter space is large and can be highly nonlinear, we start by searching for local minima throughout the full parameter space. Because the anomaly is relatively weak and smooth, we use a hot Markov-chain Monte Carlo (MCMC) as implemented in emcee (Foreman-Mackey et al. 2013) to do the search. The initial parameters are as follows: $(t_0, u_0, t_E, \rho)$ from the 1L1S fit with $3\sigma$ random scatter; $\log s$ is randomly generated from $-1 < \log s < 1$; $\alpha$ is randomly generated from $0° \leqslant \alpha < 360°$; and with the knowledge of the absence or weak central caustic crossing and central cusp perturbation of the event[8] (see Equation 11 in Chung et al. 2005), we set

$$\log q = \log(u_0/2) + 2|\log s| - \log 4. \quad (26)$$

We reduce the log-likelihood by a factor of 9 to heat the MCMC chain so that it can basically cover the whole parameter space. We adopt 200 random walkers and run for 3000 steps after 10000 steps for burn-in. After the local minima are returned from the hot chain, we perform a normal temperature MCMC on each distinct local minima to obtain the parameters and their uncertainties. Figure 7 shows the hot MCMC results in $(\log s, \log q, \alpha, \log \rho)$ space together with the refined normal MCMC over each local minima. We finally find 8 local minima in total, which are summarized in Table 4. Except for the two $q \sim 0.1$ solutions that can be easily seen in Figure 7, we zoom in on the $(\log s, \log q)$ plane in Figure 8 to show the cluster of the other local minima. The topology of the source trajectories and binary-lens caustics are shown in Figure 9.

---

[8] We take the central caustic width to be $\Delta\xi_c \approx u_0/2$, and then $\log(s - s^{-1})^2 \approx 2|\log s|$. The notations are the same as in Chung et al. (2005).

The solutions can be organized into several groups. The first group includes C1 and W1, whose central caustics and trajectories are almost identical. Their mass ratios $q$ are similar, and the separations $s$ differ by approximately $s \leftrightarrow s^{-1}$. This is the well-known "close-wide" (or the unified "inner-outer") degeneracy (Griest & Safizadeh 1998; Dominik 1999; An 2005). The source size of both solutions are precisely measured because the trajectories cross the central caustic. The second group is C2, C3, and C4. They all have $s < 1$, but with slightly different $s$ and $q$. However, their $\rho$ measurements are quite different. C3 and C4 have non-zero $\rho$ while C2 is consistent with a point source. In addition, the $\alpha$ of C3 is different from C2 and C4, corresponding to the "upper-lower" or "inner-outer" degeneracy (Gaudi & Gould 1997). We tried to change C2's $\alpha$ to the corresponding "upper" case to get a corresponding non-$\rho$ model. However, after many MCMC iterations, the solution finally converges to C3. The next group is W2 and W3. They both have $s > 1$ separations and are consistent with point source scenarios. They only differ in $\alpha$. They both have almost horizontal trajectories but do not strongly interact with the planetary caustics because there are no obvious planetary caustic crossing features in the light curve. The last group is C5. It has two local minima C5 and C5b, corresponding to the near-resonant and resonant cusics. The two local minima merge into each other because the $\chi^2$ gap between them is shallow. This solution also has a $\rho$ measurement.

From Table 4, the best-fit 2L1S model is C3, and it is $\Delta \chi^2 \sim 19$ worse than the xallarap model. The other 2L1S models (C5, W2, C4, C2, W3, C1, W1) are disfavored by $\Delta \chi^2 = (0.2, 3.0, 10.7, 15.1, 16.9, 22.8, 24.8)$, respectively. However, we find a lot of the $\chi^2$ difference between these models comes from overfitting of some low SNR features in the light curve. In Figure 10, we plot and highlight all the 2L1S and the xallarap model light curves together with the cumulative $\Delta \chi^2$ to illustrate the overfitting. For example, Model W2 has a spike at $t \sim 8732$, but there are no data points over the peak of the spike. The data points on the wings of the spike which contribute most of the $\Delta \chi^2$ are taken at the time close to the full moon, thus having small SNRs. Similar arguments can be made for Models W3, C2, C3, and C4. The $\Delta \chi^2$ of these models are mainly contributed by low-SNR data.

To separate out the influence of the low SNR data, as we mentioned in Section 4.3, in the lowest panel of Figure 10 we plot the cumulative $\Delta \chi^2$ in a time window of about $t_0 \pm 4.5$ d, corresponding to $u < 0.3$ or $I < 18.7$. This region is where the asymmetric feature occurs, and it contains most of the $\chi^2$ improvement relative to the 1L1S models. The highly magnified source flux makes the photometry more accurate in this region. We list the total $\chi^2$ of this time region $\chi^2_{\mathrm{peak}}$ in Table 4. Looking at the peak region only, the $\Delta \chi^2$s between models are reduced. The order has changed as well. For the peak region, the best-fit model is now C5, and then (C4, C3, C2, W2, W3, C1, W1) are disfavored by $\Delta \chi^2 = (0.5, 0.8, 4.5, 8.1, 14.0, 16.0, 20.2)$, respectively. The $\chi^2$ difference between the 2L1S models and the xallarap models are reduced as well.

We also tried testing parallax effects in 2L1S models, but neither significant improvements nor useful constraints were obtained. Therefore, we do not include parallax in the final 2L1S models.

### 4.5 Summary of Light-Curve Analysis

After the exploration, we exclude the 1L1S, 1L1S with parallax, and 1L2S scenarios for MOA-2019-BLG-421. The remaining models are the 1L1S with xallarap and the 2L1S models. For the 2L1S interpretations, there are many degenerate models that can explain the light curve almost equally well. However, as will be shown in the Sections 5 and 6, the solutions with finite source $\rho$ measurements are very unlikely to be correct.

## 5 SOURCE PROPERTIES OF MOA-2019-BLG-421

The source star color can be used to measure the angular radius of the source star, $\theta_*$. With the source radius, the angular Einstein radius and the relative proper motion can be obtained,

$$\theta_{\mathrm{E}} = \frac{\theta_*}{\rho}, \quad \mu_{\mathrm{rel}} = \frac{\theta_{\mathrm{E}}}{t_{\mathrm{E}}}, \tag{27}$$

which are directly related to the physical parameters of the lens.

To measure the color of the source star, first, we construct a Color-Magnitude Diagram (CMD) from stars within a $2' \times 2'$ square centered on the source position using KMTC03 images. The magnitude and color are calibrated to the OGLE-III catalog (Szymański et al. 2011). The CMD is shown in Figure 11.

Then, we place the microlensing source on the CMD. We determine the source $I$-band magnitude from the models (Tables 3 and 4) and the color $(V - I) = 1.569 \pm 0.073$ from the linear regression of the $V$-band and $I$-band source fluxes during the event. The color and magnitude are also calibrated to OGLE-III.

Next, we measure the centroid of the red clump giants (following the method in Nataf et al. (2013)) to be $(V-I)_{\mathrm{RC}} = 1.787 \pm 0.016$ and $I_{\mathrm{RC}} = 15.313 \pm 0.054$. We measure the offset of the source relative to the centroid of the red clump giants (Yoo et al. 2004). By comparing the instrumental $[(V - I), I]_{\mathrm{RC}}$ with the intrinsic centroid of the red giant clump $[(V - I), I]_{\mathrm{RC},0} = [1.06, 14.339]$ from Bensby et al. (2013) and Nataf et al. (2013), we can find the intrinsic, de-reddened color and magnitude of the source.

Finally, based on the de-reddened color and magnitude, we estimate the source angular radius $\theta_*$ from Adams et al. (2018). The de-reddened source colors and magnitudes together with the derived $(\theta_*, \theta_{\mathrm{E}}, \mu_{\mathrm{rel}})$ of all 2L1S solutions and the xallarap solutions are listed in Table 5.

We immediately see that for the solutions with finite source measurement, the derived $\theta_{\mathrm{E}}$ and $\mu_{\mathrm{rel}}$ are unusual. For example, if we consider the case that the source and the lens are both in the Galactic bulge, we would expect $\mu_{\mathrm{rel}} > (0.93, 2.44)$ for $(3\sigma, 2\sigma)$ limits, respectively. The limits are not very sensitive to the Galactic components because the probability for a very small $\mu_{\mathrm{rel}}$ lens-source pair to create microlensing events is small. Therefore, Solutions (C1, W1, C3, C4, C5) are unlikely to be real. However, if we consider the detection bias introduced by detected planetary events, the expected $\mu_{\mathrm{rel}}$ distribution would be systematically moved toward small values (Gould 2022) because longer planetary perturbations are easier to detect. The $\mu_{\mathrm{rel}}$ limits are then $> (0.32, 1.30)$ for $(3\sigma, 2\sigma)$. The Solutions (C1, W1, C3, C4, C5) are still disfavored but with less confidence.

## 6 LENS PROPERTIES OF MOA-2019-BLG-421

To uniquely obtain the physical parameters of the lens system $D_{\mathrm{L}}$ and $M_{\mathrm{L}}$, one needs at least two of $\theta_{\mathrm{E}}$, $\pi_{\mathrm{E}}$, and the absolute brightness of the lens object (see also, e.g., Introduction of Zhang et al. 2020). However, for MOA-2019-BLG-421, we only have $\theta_{\mathrm{E}}$ or its lower limit. Therefore we perform a Bayesian analysis to obtain the posterior distribution of the physical parameters of the lens. The prior of the Bayesian analysis is the Galactic model, including the stellar density profile, the mass function, and the stellar velocity distribution. We adopt "Model C" described in Yang et al. (2021).

We generate $10^8$ microlensing events based on the Galactic model, that is, generating the source and lens distance from the line-of-sight stellar density profiles, lens mass from the mass function, and source and lens motions from the stellar velocity distribution. The underlying assumption is that the probability of a star hosting a planet is independent of its mass and Galactic environment. For each simulated event, $i$, we weight it by

$$w_i = \Gamma_i \times \mathcal{L}_i(t_{\rm E}) \mathcal{L}_i(\theta_{\rm E}), \tag{28}$$

where $\Gamma_i \propto \theta_{\rm E,i} \mu_{\rm rel,i}$ is the microlensing event rate. $\mathcal{L}(t_{\rm E})$ and $\mathcal{L}(\theta_{\rm E})$ are the likelihood function of $t_{\rm E}$ and $\theta_{\rm E}$ measured for a specific solution from Section 4. The summation $\sum_i w_i$ is proportional to the total event rate $\Gamma_{\rm tot}$ for a specific solution. We use a Gaussian likelihood function for $t_{\rm E}$, and use a Gaussian likelihood for $\theta_{\rm E}$ if it is measured or a flat distribution with $3 - \sigma$ lower limit hard cut if the value is not measured. To consider the $\mu_{\rm rel}$ detection bias proposed by Gould (2022), we also report the total event rate $\Gamma'_{\rm tot}$ using the weight $w'_i = w_i/\mu_{\rm rel,i}$.

The final results of the physical parameters of all models are shown in Table 6. For the xallarap models, the lens is an M dwarf located in the Galactic bulge. For the 2L1S models, if Model (C1, W1) is correct, the lens system is likely to be a brown dwarf orbited by a super Jupiter and their projected separation is $\sim (0.05, 1.06)$ AU. In the case of Model (C3, C4, C5), the lens system is a super Earth or Neptune orbiting a low-mass M dwarf at a projected separation of $\sim (0.15, 0.30, 0.16)$ AU. For Model (C2, W2, W3), the lens system is an M dwarf of low-mass K dwarf with a sub-Jupiter mass planet. The projected separations are $\sim (1.4, 2.6, 2.6)$ AU, respectively.

We list the relative event rate of all models in Table 6. It is hard to generate events like Model (C1, W1, C3, C4, C5) in the Galaxy compared to Model (C2, W2, W3). Therefore, the former models are strongly disfavored under the Galactic prior, a result that is consistent with our preliminary discussion in Section 5. We convert the relative event rate in terms of $\chi^2$ using

$$\chi^2_{\rm Gal.+\mu_{\rm rel}} = -2 \ln \Gamma'_{\rm tot}. \tag{29}$$

Then we combine $\chi^2_{\rm peak}$ in Table 4 with the $\chi^2_{\rm Gal.+\mu_{\rm rel}}$, to obtain the results listed in Table 6. Finally, we conclude that Models C2 and W2 are preferred among all the 2L1S interpretations.

This preference can also be tested by future observations measuring the relative proper motion $\mu_{\rm rel}$ and/or the lens light. The different $\mu_{\rm rel}$ predictions for different models are already shown in Tables 5 and 6. For the lens light, if the lens is a main-sequence star located in the Galactic bulge, the brightness would be $(I_{\rm L}, K_{\rm L}) = (22.9^{+3.4}_{-2.4}, 20.7^{+2.0}_{-1.4})$ for Models (C2, W2, W3) or $(I_{\rm L}, K_{\rm L}) = (30.7^{+3.1}_{-3.5}, 25.5^{+1.8}_{-3.5})$ for Models (C1, W1, C3, C4, C5). If the currently preferred models (C2, W2, W3) are correct, assuming the $\mu_{\rm rel} \sim 7 \pm 3$ mas/yr from Table 6, the lens and the source will be separated by $\Delta\theta \sim 70$ mas in 2030. Given the similar brightness of the lens and source ($I_{\rm S} \sim 20.2$, $K_{\rm S} \sim 18.2$), a measurement of the lens light would be achievable on the current largest telescopes (e.g., Keck, VLT). However, if the other models (C1, W1, C3, C4, C5) are correct, measurement of $\mu_{\rm rel}$ will be challenging even with the next-generation telescopes given the small $\mu_{\rm rel}$ and large contrast ratios.

In addition, if the xallarap model is correct, we expect a large radial velocity for the source star ( RV amplitude $K \sim 30$ km s$^{-1}$ assuming $M_{\rm S} = 1 M_\odot$, $q_{\rm S} = 0.5$, and 60° inclination). However, as a result of the faint brightness $I_{\rm S} \sim 20$, the radial velocity measurement of the source would be very challenging. Therefore, technically, it will be hard to resolve the degeneracy between the best 2L1S models and the xallarap models.

# 7 IMPROVEMENTS

We updated the KMTNet TLC photometry procedures to increase the automation and reduce the need for highly-skilled operators, as well as making a few modifications to increase the photometric accuracy. With our new pipeline, there are a total of three different versions of pySIS for reducing KMTNet data: the preliminary pySIS pipeline (also called online pipeline or end-of-year pipeline), the by-hand TLC procedure, and the new TLC pipeline from this work. Here we compare these versions and discuss the improvements.

## 7.1 Efficiency Improvements

For efficiency, here we compare the by-hand TLC and the new TLC pipeline, because we aim to reduce the time cost of TLC so that it can be used to search for new anomaly signals.

The new TLC pipeline is more parallelized, which makes it perform better on multi-core machines. The original pySIS only parallelized image subtractions because image subtraction is the most computationally expensive step. This step can be sped up by increasing the number of CPU cores, at which point the other steps that were not parallelized, such as alignment and photometry, become the bottlenecks. Therefore, we parallelized all steps to improve multi-core performance.

The major improvement is that the new TLC pipeline is more automated. Without that automation, human operators had to wait for preprocessing to finish, then select reference images by eye. After that, the pipeline would align all the images and create the master reference image. The operator was then required to enter the target position. For KMTNet data, this setup process took about ~1 hour. Although it did not require active participation by the operator the entire time, it did require the operator to check its status frequently during this period. Then, the remainder of the pipeline would run without operator input; a process that took ~ 1 to 2 hours for KMT-Net data on a single core. If the results were not ideal, the operator would have to start from the beginning.

The human components of the by-hand TLC procedure made the results highly operator-dependent. Because the original pySIS had limited quantitative parameters to describe the qualities of the reference images and other results, robust reference image selection required either extensive experience, external algorithms, luck, or a combination of all three. For some operators, the best approach was to do several iterations with different sets of reference images to test what combinations worked best.

The automations in the new TLC pipeline remove intermediate human interactions, significantly reducing the human workload in executing the reduction. In addition, because new metrics help automatically and robustly select reference images, in most cases, the photometry is good quality on the first try, so the reduction procedure usually does not need to be repeated multiple times. We also added functionality to the code to allow operators to re-start the process from any intermediate step if further adjustments are needed.

For the by-hand TLC reductions, the typical time cost for a prime-field event (~ 6 × 3000 images) is 6-8 hours operating on a 50-core machine, of which 1-3 hr required some level of human attention. Therefore, if we were to systematically run the TLC pipeline for all prime-field events during one season (~ 1000), a total of ~ 300 days would be needed, including significant amounts of human review and potentially additional iterations. After the updates, for a prime-field event, the typical computational time cost is now reduced to ~ 1 hr plus 10 – 30 min for a manual check in the end. These automations and additional parallelizations make systematic reanalysis possible.

### 7.2 Accuracy Improvements

We performed a series of tests to better understand what aspects of the new TLC pipeline most affect the photometry for MOA-2019-BLG-421. We reduced the data using the old TLC pySIS procedure without any special optimizations, e.g., the reference images were selected by eye based on a list sorted by FWHM. Then, we individually fit the various datasets to a PSPL model using the raw errorbars.

For these fits, we calculated the mean, median, and standard deviations of the $\chi^2$ contribution from each datapoint to use as metrics for assessing the quality of each dataset. We also calculated the mean, median, and standard deviation of the absolute value of the residuals after scaling the fluxes to the same system. See Table 7. From the residuals, we can see that data from the new TLC pipeline has much less scatter than either the online or by-hand TLC reductions. For the $\chi^2$ per point, we expect a value of 1 for Gaussian statistics. So, we see that the errorbar estimation from the by-hand TLC data is better than for the online data and that the new TLC pipeline may slightly over-estimate the errorbars in this case.

We also tried repeating the by-hand TLC procedure to test some of the changes in the new TLC pipeline. For example, we tried using the same reference images and lens position in the by-hand TLC reduction as for the new TLC pipeline. These changes did not have a significant effect on the quality of the photometry. Hence, we can conclude that, in this case, the photometry improvement from the new TLC pipeline is due to some other optimization.

Of course, the pySIS algorithms were never intended to produce optimal photometry in all cases without any optimizations, and our comparison to the un-optimized by-hand TLC reduction does not clearly distinguish between improvements to the photometry due to improved photometry algorithms and those due better choices of default parameters (i.e., tuned for KMTNet datasets). So, it is still possible that an expert operator could produce photometry of similar quality to the new TLC pipeline. However, this requires deep, specialized knowledge of the underlying algorithms and photometry parameters. On the other hand, for this particular dataset, most of the significantly magnified points are concentrated in a few nights around the peak, which happen to have seeing in the 10th percentile, meaning it is often better than the reference images. So, the improvements that reduce seeing correlations in the data may also be significant in this case. Regardless, this test demonstrates that the new TLC pipeline can produce much better quality photometry in this case without a lot of effort on the part of the operator.

### 7.3 Comparison to Online Data

In addition, the key that powers the new anomaly search are the improvements from the online data to the new TLC data.

To quantify this improvement, we follow the procedure in Yang et al. (2022) to calculate the planetary sensitivity of MOA-2019-BLG-421, using the online data and the new TLC data, respectively. To quantify the difference between the two data sets, we calculate the planet sensitivity for each one following the procedure of Rhie et al. (2000) as described in Yang et al. (2022). In short, we generate a series of artificial 2L1S light curves using the actual noise from the real light curves. For each artificial light curve, we find its deviations to the 1L1S model. Therefore, the $\chi^2$ difference between the 2L1S and 1L1S models represents the significance of the artificial signal. Figure 12 shows the $\chi^2$ distribution on $(s, \alpha)$ plane for an injected log $q = -2.8$ planet. Each point represents an artificial light curve generated by the given $(s, q, \alpha)$ and the color represents the $\Delta\chi^2$. We define $\Delta\chi^2 > 100$ for a detection. The rightmost panel of Figure 12 shows the "detection" region enhanced by the new TLC data.

We marginalize the $\Delta\chi^2 > 100$ probability over $\alpha$ to obtain the sensitivity over $(s, q)$ plane. The results are shown in Figure 13. For the actual planetary signals detected in MOA-2019-BLG-421, the sensitivity changes from $< 20\%$ in the online data to $> 80\%$ with the new TLC photometry. For the full $(-0.3 \leqslant \log s \leqslant 0.3, -4.0 \leqslant \log q \leqslant -2.3)$ phase space region, the sensitivity is improved from ~ 24% to ~ 53%.

In conclusion, the new TLC data can significantly improve the detection sensitivity of planets (or planetary-like anomalies). In the specific case of MOA-2019-BLG-421, the new TLC increases the $\Delta\chi^2$ for the anomaly above the threshold, which is why the new anomaly could be detected.

However, new TLC data would not significantly change our knowledge about the planets that have been published. In order for a planet to be published in the first place, the signal needs to be clear, so any improvements would simply tend to improve the clarity of known signals. However, the accuracy improvements from the new TLC pipeline relative to the online data do allow us to find previously undiscovered planets or clarify previously unpublishable signals.

## 8 DISCUSSION

In this work, we updated the KMTNet TLC photometry algorithms to improve their automation and photometric accuracy. By applying the new TLC pipeline to historic events in the KMTNet database, we find a new anomaly signal in MOA-2019-BLG-421, which was buried in the noise of the preliminary data. The signal can be explained by either the orbital motion of the source star or a planet in the lens system. For the planetary interpretation, the planetary system is most likely to be a Jovian planet orbiting an M dwarf in the Galactic bulge, which is a typical microlensing planetary system. The discovery shows that there are indeed missed signals under current planet search procedures. These updated photometric data can indeed increase the sensitivity of the KMTNet survey.

Apart from the accuracy improvements, the new TLC pipeline automates human interactions in the middle of the TLC reduction procedure both decreasing the human workload and making the results more robust. Together with some minor improvements regarding parallelization, the typical time cost for reducing a KMTNet 15,000 image prime-field event is reduced to ~ 1 hr using a 64 CPU-core device. The remaining human work (check the final results) is reduced to the order of minutes. The overall time cost is now ≲ 10% of the previous, by-hand TLC procedure. With such improvements in efficiency and automation, a systematic search of hundreds of events becomes possible.

Although the optimized data can potentially enable us to obtain better statistical results, a statistical sample must be defined carefully. Currently, although the new TLC pipeline reduced many of the human efforts, a human reviewer is still needed to verify the final results and deal with special cases. Applying it to all KMTNet events of the past 7 seasons (~ 18,000 events in total) is difficult. There are two feasible options. The first is to run the pipeline without any human reviews and exclude bad data though the quality indicators. Another is to define a sub-sample on the order of hundreds of events and apply the full pipeline including human reviews. As discussed in Yee et al. (2021) and Zang et al. (2021a), one approach is to compile a sample of high-magnification events, because they are intrinsically more sensitive to planets. A small number of such events could contribute a large fraction of the total survey sensitivity. Another approach could be compiling a sample with a specific type of source stars, e.g., giant sources. These sources are also intrinsically more sensitive to planetary perturbations than average. In addition, the higher luminosity of such sources could significantly increase the photometric accuracy, especially for the data reduced with the new TLC pipeline. We will implement these approaches in the near future for this systematic re-analysis project.

## ACKNOWLEDGEMENTS

This research has made use of the KMTNet system operated by the Korea Astronomy and Space Science Institute (KASI) at three host sites of CTIO in Chile, SAAO in South Africa, and SSO in Australia. Data transfer from the host site to KASI was supported by the Korea Research Environment Open NETwork (KREONET). H. Yang, Q. Qian, J. Zhang, and S. Mao acknowledge support from the National Natural Science Foundation of China (Grant No. 12133005). Z. Hu and W. Zhu were also supported by the National Science Foundation of China (grant No. 12173021). J.C.Y., I.-G.S., and S.-J.C. acknowledge the support from NSF Grant No. AST-2108414. W.Zang acknowledges the support from the Harvard-Smithsonian Center for Astrophysics through the CfA Fellowship. The authors acknowledge the Tsinghua Astrophysics High-Performance Computing platform at Tsinghua University for providing computational and data storage resources that have contributed to the research results reported within this paper.

## DATA AVAILABILITY

**Table 1.** Published events using the updated photometry data

| Event Name | KMT Name | Reference |
|---|---|---|
| OGLE-2016-BLG-1635 | KMT-2016-BLG-0269 | Shin et al. (2023a) |
| OGLE-2016-BLG-1195 | KMT-2016-BLG-0372 | Gould et al. (2023a) |
| MOA-2016-BLG-532 | KMT-2016-BLG-0506 | Shin et al. (2023a) |
| KMT-2016-BLG-1105 | KMT-2016-BLG-1105 | Zang et al. (2023) |
| KMT-2016-BLG-1751 | KMT-2016-BLG-1751 | Shin et al. (2023a) |
| KMT-2016-BLG-1855 | KMT-2016-BLG-1855 | Shin et al. (2023a) |
| KMT-2017-BLG-0428 | KMT-2017-BLG-0428 | Zang et al. (2023) |
| KMT-2017-BLG-1003 | KMT-2017-BLG-1003 | Zang et al. (2023) |
| OGLE-2017-BLG-1806 | KMT-2017-BLG-1021 | Zang et al. (2023) |
| KMT-2017-BLG-1194 | KMT-2017-BLG-1194 | Zang et al. (2023) |
| OGLE-2019-BLG-0249 | KMT-2019-BLG-0109 | Jung et al. (2023) |
| KMT-2019-BLG-1367 | KMT-2019-BLG-1367 | Zang et al. (2023) |
| KMT-2019-BLG-1806 | KMT-2019-BLG-1806 | Zang et al. (2023) |
| OGLE-2019-BLG-0679 | KMT-2019-BLG-2688 | Jung et al. (2023) |
| KMT-2021-BLG-0119 | KMT-2021-BLG-0119 | Shin et al. (2023b) |
| KMT-2021-BLG-0192 | KMT-2021-BLG-0192 | Shin et al. (2023b) |
| KMT-2021-BLG-2294 | KMT-2021-BLG-2294 | Shin et al. (2023b) |
| KMT-2022-BLG-0371 | KMT-2022-BLG-0371 | Han et al. (2023b) |
| KMT-2022-BLG-0440 | KMT-2022-BLG-0440 | Zhang et al. (2023) |
| KMT-2022-BLG-1013 | KMT-2022-BLG-1013 | Han et al. (2023b) |
| KMT-2022-BLG-2397 | KMT-2022-BLG-2397 | Gould et al. (2023b) |

This paper has been typeset from a TEX/LATEX file prepared by the author.

**Table 2.** Parameters of static 1L1S, 1L1S + parallax, and 1L2S solutions for MOA-2019-BLG-421

| Model | $t_0$ (HJD′) | $u_0$ | $t_E$ (d) | $\pi_{E,N}$ | $\pi_{E,E}$ | | $I_s$ | $\chi^2$/dof |
|---|---|---|---|---|---|---|---|---|
| 1L1S | 8743.0705 | 0.0222 | 15.23 | – | – | | 20.17 | 6962.4/6792 |
| | 0.0011 | 0.0009 | 0.28 | | | | 0.06 | |
| 1L1S parallax | 8743.0716 | 0.0199 | 16.31 | −11.77 | 2.40 | | 20.19 | 6888.9/6790 |
| ($u_0 > 0$) | 0.0011 | 0.0007 | 0.48 | 1.20 | 0.29 | | 0.06 | |
| 1L1S parallax | 8743.0714 | −0.0199 | 16.60 | −11.79 | 2.45 | | 20.19 | 6888.9/6790 |
| ($u_0 < 0$) | 0.0011 | 0.0007 | 0.53 | 1.23 | 0.29 | | 0.06 | |

| Model | $t_{0,1}$ (HJD′) | $u_{0,1}$ | $t_E$ (d) | $t_{0,2}$ (HJD′) | $u_{0,2}$ | $q_{F,I}$ | $q_{F,\text{MOA}-R}$ | $I_{s,1}$ | $\chi^2$/dof |
|---|---|---|---|---|---|---|---|---|---|
| 1L2S | 8743.0729 | 0.0214 | 14.92 | 8741.95 | 0.065 | 0.047 | < 0.117 | 20.19 | 6857.8/6788 |
| | 0.0015 | 0.0007 | 0.34 | 0.15 | 0.012 | 0.012 | – | 0.06 | |

NOTE. HJD′=HJD−2450000. The parameters and their $1\sigma$ uncertainties are presented. For the non-detection parameters, the $3\sigma$ upper limits are provided. No useful $\rho$ is measured in these models.

**Table 3.** Parameters of 1L1S xallarap solutions for MOA-2019-BLG-421

| Model | $t_0$ | $u_0$ | $t_E$ (d) | $P_\xi$ (d) | $\phi_\xi$ | $i_\xi$ | $\xi_{E,N}$ | $\xi_{E,E}$ | $I_s$ | $\chi^2$/dof | $\chi^2_{\text{peak}}$/dof |
|---|---|---|---|---|---|---|---|---|---|---|---|
| 1L1S | 8743.0756 | 0.0276 | 12.32 | 14.15 | 3.59 | 1.38 | 0.0193 | −0.0451 | 19.83 | 6772.9/6787 | 419.6/431 |
| XLRP+ | 0.0012 | 0.0018 | 0.79 | 1.33 | 0.31 | 0.22 | 0.0078 | 0.0123 | 0.09 | | |
| 1L1S | 8743.0758 | 0.0270 | 12.11 | 14.29 | 0.45 | 1.82 | −0.0203 | −0.0573 | 19.86 | 6769.4/6787 | 418.6/431 |
| XLRP− | 0.0011 | 0.0018 | 0.74 | 1.06 | 0.24 | 0.09 | 0.0074 | 0.0122 | 0.09 | | |

NOTE. HJD′=HJD−2450000. "XLRP" represents "xallarap". The peak region is defined by $t_0 \pm 4.5$ d.

**Table 4.** Parameters of 2L1S solutions for MOA-2019-BLG-421

| Model | $t_0$ (HJD′) | $u_0$ | $t_E$ (d) | $\rho$ ($10^{-3}$) | $\alpha$ | $s$ | $q$ ($10^{-4}$) | $I_s$ | $\chi^2$/dof | $\chi^2_{\text{peak}}$/dof |
|---|---|---|---|---|---|---|---|---|---|---|
| C1 | 8743.0422 | 0.0252 | 15.39 | 22.61 | 1.072 | 0.245 | 1243.9 | 20.20 | 6810.5/6788 | 433.3/432 |
| | 0.0030 | 0.0007 | 0.31 | 0.82 | 0.028 | 0.009 | 142.7 | 0.06 | | |
| W1 | 8743.0497 | 0.0228 | 17.13 | 20.09 | 1.072 | 5.010 | 1732.3 | 20.23 | 6812.5/6788 | 437.4/432 |
| | 0.0025 | 0.0005 | 0.35 | 0.60 | 0.030 | 0.030 | 273.2 | 0.06 | | |
| **C2** | **8743.0449** | **0.0198** | **16.26** | **< 9.52** | **3.199** | **0.700** | **14.5** | **20.26** | **6802.8/6788** | **421.7/432** |
| | **0.0021** | **0.0005** | **0.32** | **–** | **0.004** | **0.006** | **1.1** | **0.06** | | |
| C3 | 8743.0642 | 0.0259 | 15.45 | 24.46 | 3.011 | 0.814 | 6.4 | 20.16 | 6787.7/6788 | 418.0/432 |
| | 0.0015 | 0.0009 | 0.32 | 1.80 | 0.006 | 0.006 | 1.0 | 0.06 | | |
| C4 | 8743.0408 | 0.0215 | 15.66 | 12.05 | 3.193 | 0.761 | 13.7 | 20.22 | 6798.4/6788 | 417.8/432 |
| | 0.0027 | 0.0005 | 0.30 | 2.09 | 0.004 | 0.005 | 1.0 | 0.06 | | |
| **W2** | **8743.0519** | **0.0230** | **14.24** | **< 11.53** | **3.202** | **1.433** | **12.1** | **20.10** | **6790.8/6788** | **425.3/432** |
| | **0.0020** | **0.0005** | **0.29** | **–** | **0.004** | **0.016** | **1.4** | **0.06** | | |
| W3 | 8743.0472 | 0.0237 | 14.13 | < 11.35 | 3.140 | 1.425 | 15.1 | 20.09 | 6804.7/6788 | 431.2/432 |
| | 0.0023 | 0.0005 | 0.28 | – | 0.005 | 0.019 | 1.7 | 0.06 | | |
| C5 | 8743.0662 | 0.0274 | 15.11 | 26.43 | 3.028 | 0.944 | 3.6 | 20.13 | 6787.9/6788 | 417.2/432 |
| | 0.0039 | 0.0009 | 0.32 | 1.19 | 0.010 | 0.021 | 0.8 | 0.06 | | |

NOTE. HJD′=HJD−2450000. The parameters and their $1\sigma$ uncertainties are presented. For the non-detection parameters, the $3\sigma$ upper limits are provided. The peak region is defined by $t_0 \pm 4.5$ d. The final preferred models (in Section 6) are highlighted in boldface.

**Table 5.** Source properties for MOA-2019-BLG-421

| Model | $(V-I)_{s,0}$ | $I_{s,0}$ | $\theta_*$ ($\mu$as) | $\theta_E$ (mas) | $\mu_{rel}$ (mas/yr) |
|---|---|---|---|---|---|
| C1 | $0.84 \pm 0.08$ | $19.22 \pm 0.08$ | $0.518 \pm 0.053$ | $0.023 \pm 0.002$ | $0.54 \pm 0.18$ |
| W1 | $0.84 \pm 0.08$ | $19.25 \pm 0.08$ | $0.512 \pm 0.052$ | $0.025 \pm 0.003$ | $0.54 \pm 0.20$ |
| **C2** | $\mathbf{0.84 \pm 0.08}$ | $\mathbf{19.29 \pm 0.08}$ | $\mathbf{0.504 \pm 0.052}$ | $\mathbf{> 0.037}$ | $\mathbf{> 0.82}$ |
| C3 | $0.84 \pm 0.08$ | $19.19 \pm 0.08$ | $0.527 \pm 0.054$ | $0.022 \pm 0.002$ | $0.51 \pm 0.17$ |
| C4 | $0.84 \pm 0.08$ | $19.24 \pm 0.08$ | $0.514 \pm 0.053$ | $0.043 \pm 0.004$ | $0.99 \pm 0.31$ |
| **W2** | $\mathbf{0.84 \pm 0.08}$ | $\mathbf{19.13 \pm 0.08}$ | $\mathbf{0.541 \pm 0.055}$ | $\mathbf{> 0.033}$ | $\mathbf{> 0.83}$ |
| W3 | $0.84 \pm 0.08$ | $19.11 \pm 0.08$ | $0.546 \pm 0.056$ | $> 0.033$ | $> 0.86$ |
| C5 | $0.84 \pm 0.08$ | $19.15 \pm 0.08$ | $0.535 \pm 0.055$ | $0.020 \pm 0.002$ | $0.49 \pm 0.17$ |
| **1L1S XLRP+** | $\mathbf{0.84 \pm 0.08}$ | $\mathbf{18.86 \pm 0.10}$ | $\mathbf{0.612 \pm 0.065}$ | $\mathbf{> 0.027}$ | $\mathbf{> 0.82}$ |
| **1L1S XLRP−** | $\mathbf{0.84 \pm 0.08}$ | $\mathbf{18.88 \pm 0.10}$ | $\mathbf{0.606 \pm 0.065}$ | $\mathbf{> 0.028}$ | $\mathbf{> 0.84}$ |

NOTE. The parameters and their $1\sigma$ uncertainties are presented. For the models without finite source effect detections, the $3\sigma$ lower limits of $\theta_E$ and $\mu_{rel}$ are provided. "XLRP" represents xallarap". The final preferred models (in Section 6) are highlighted in boldface.

**Table 6.** Physical parameters of the lens (system) from Bayesian analysis for MOA-2019-BLG-421

| Model | $M_{host}$ ($M_\odot$) | $M_P$ ($M_J$) | $D_L$ (kpc) | $D_{LS}$ (kpc) | $a_\perp$ (AU) | $\mu_{rel}$ (mas/yr) | Relative $\Gamma_{tot}$ | $\Gamma'_{tot}$ | $\chi^2_{Gal.+\mu_{rel}}$ | $\chi^2_{peak} + \chi^2_{Gal.+\mu_{rel}}$ |
|---|---|---|---|---|---|---|---|---|---|---|
| C1 | $0.039^{+0.067}_{-0.021}$ | $5.1^{+9.0}_{-2.8}$ | $7.9^{+0.7}_{-0.7}$ | $0.11^{+0.13}_{-0.07}$ | $0.047^{+0.008}_{-0.006}$ | $0.55^{+0.05}_{-0.06}$ | 1.22 | 1.17 | −0.3 | 432.9 |
| W1 | $0.039^{+0.076}_{-0.021}$ | $7.0^{+14.0}_{-3.9}$ | $8.0^{+0.7}_{-0.7}$ | $0.13^{+0.14}_{-0.09}$ | $1.06^{+0.17}_{-0.15}$ | $0.52^{+0.06}_{-0.05}$ | 1.24 | 1.23 | −0.4 | 437.0 |
| **C2** | $\mathbf{0.41^{+0.38}_{-0.25}}$ | $\mathbf{0.62^{+0.58}_{-0.39}}$ | $\mathbf{6.9^{+0.8}_{-1.1}}$ | $\mathbf{1.56^{+1.41}_{-0.75}}$ | $\mathbf{1.37^{+0.52}_{-0.48}}$ | $\mathbf{6.6^{+2.8}_{-2.4}}$ | **58906.97** | **5243.76** | **−17.1** | **404.5** |
| C3 | $0.039^{+0.068}_{-0.021}$ | $0.026^{+0.046}_{-0.014}$ | $7.9^{+0.7}_{-0.7}$ | $0.11^{+0.14}_{-0.07}$ | $0.15^{+0.03}_{-0.02}$ | $0.55^{+0.07}_{-0.07}$ | 1.24 | 1.17 | −0.3 | 417.7 |
| C4 | $0.054^{+0.085}_{-0.032}$ | $0.076^{+0.123}_{-0.044}$ | $7.7^{+0.7}_{-0.7}$ | $0.42^{+0.55}_{-0.27}$ | $0.30^{+0.09}_{-0.06}$ | $1.21^{+0.35}_{-0.24}$ | 39.89 | 17.18 | −5.7 | 412.1 |
| **W2** | $\mathbf{0.37^{+0.36}_{-0.23}}$ | $\mathbf{0.46^{+0.47}_{-0.29}}$ | $\mathbf{7.0^{+0.7}_{-1.0}}$ | $\mathbf{1.45^{+1.30}_{-0.71}}$ | $\mathbf{2.58^{+0.99}_{-0.89}}$ | $\mathbf{6.9^{+2.9}_{-2.4}}$ | **81719.15** | **6955.51** | **−17.7** | **407.6** |
| W3 | $0.37^{+0.36}_{-0.23}$ | $0.58^{+0.59}_{-0.36}$ | $7.0^{+0.7}_{-1.0}$ | $1.44^{+1.29}_{-0.70}$ | $2.55^{+0.99}_{-0.88}$ | $6.9^{+2.9}_{-2.5}$ | 80423.49 | 6823.77 | −17.7 | 413.6 |
| C5 | $0.039^{+0.065}_{-0.021}$ | $0.015^{+0.025}_{-0.009}$ | $8.0^{+0.7}_{-0.7}$ | $0.09^{+0.12}_{-0.06}$ | $0.16^{+0.03}_{-0.02}$ | $0.52^{+0.06}_{-0.05}$ | 1.00 | 1.00 | −0.0 | 417.2 |
| **1L1S XLRP** | $\mathbf{0.30^{+0.35}_{-0.20}}$ | – | $\mathbf{7.0^{+0.7}_{-0.9}}$ | $\mathbf{1.35^{+1.21}_{-0.67}}$ | – | $\mathbf{7.3^{+3.0}_{-2.6}}$ | | | | |

NOTE. The median value and the $\pm 1\sigma$ range of the posterior distribution of the parameters are presented. $D_{LS} = D_S - D_L$. The relative $\Gamma_{tot}$ is the event rate corresponding to each model. The relative $\Delta\chi^2$ introduced by the Galactic model and the $\mu_{rel}$ bias is defined as $\Delta\chi^2_{Gal.+\mu_{rel}} = -2\ln\Gamma'_{tot}$. The $\chi^2_{peak}$ is adopted from Tables 3 and 4. The final preferred models are highlighted in boldface.

**Table 7.** Photometry Quality Metrics for KMTS03

| All Data | $\chi^2$/pt | | | | Residuals (mag) | | |
|---|---|---|---|---|---|---|---|
| dataset | N | Mean | Med | Std | Sum | Mean | Med | Std |
| online | 1870 | 10.69 | 4.47 | 33.44 | 19990.12 | 0.310 | 0.166 | 0.432 |
| by-hand TLC | 2435 | 6.48 | 1.58 | 31.63 | 15773.26 | 0.392 | 0.236 | 0.581 |
| new pipeline | 1708 | 1.35 | 0.33 | 3.59 | 2298.69 | 0.140 | 0.084 | 0.219 |
| **Peak Data** | $\chi^2$/pt | | | | Residuals (mag) | | |
| dataset | N | Mean | Med | Std | Sum | Mean | Med | Std |
| online | 55 | 5.98 | 2.43 | 8.50 | 329.02 | 0.056 | 0.025 | 0.084 |
| by-hand TLC | 57 | 3.67 | 1.62 | 5.01 | 208.96 | 0.066 | 0.035 | 0.100 |
| new pipeline | 50 | 0.75 | 0.22 | 1.52 | 37.73 | 0.025 | 0.014 | 0.029 |

"Peak" metrics are calculated from points within $t_0 \pm 2.5$ d.

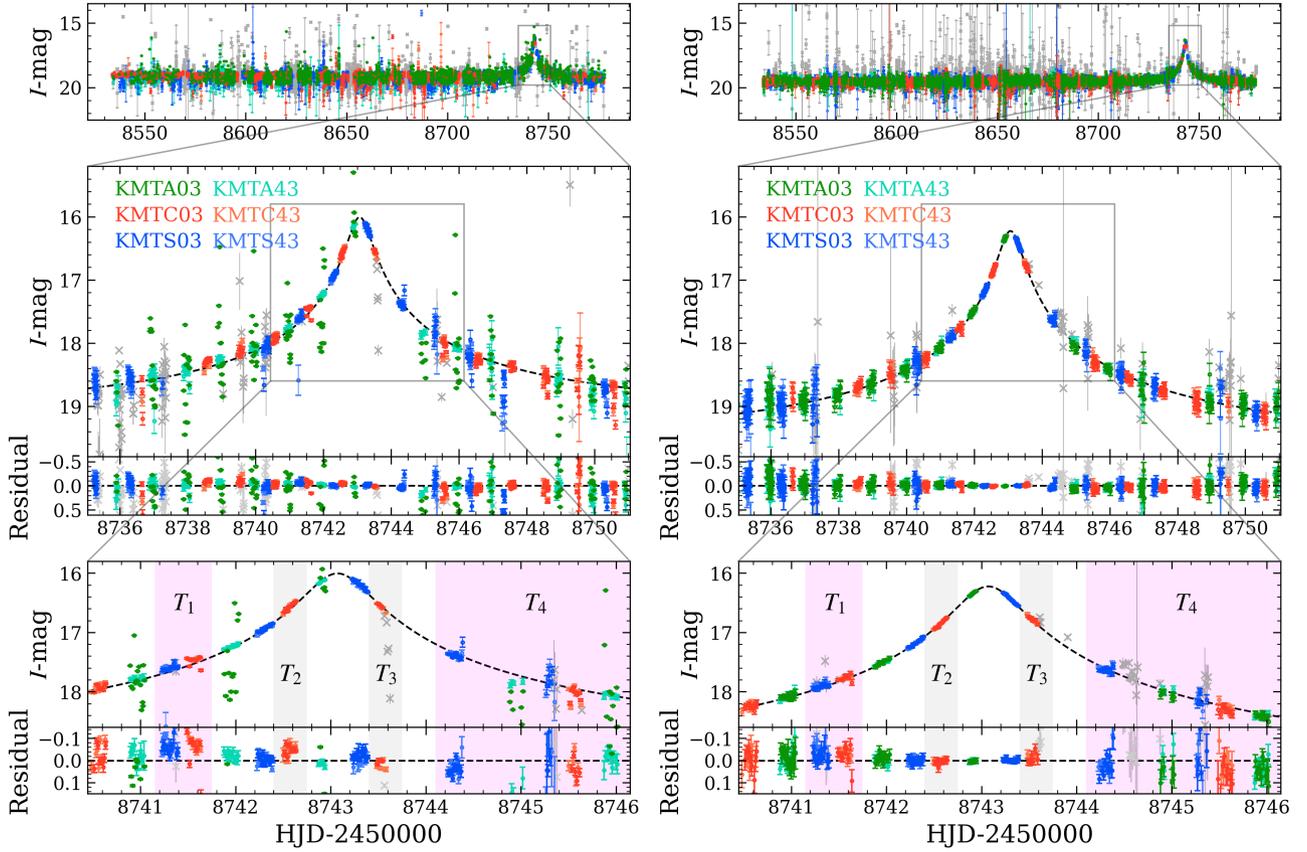

**Figure 4.** Light curves of MOA-2019-BLG-421. *(left).* The preliminary data and the best PSPL fit to them. *(right).* The TLC data from the updated algorithm and the best PSPL fit to them. The error bars are the native error bars from the photometry pipeline. The gray "x" points in the left panels are excluded by the AnomalyFinder algorithm (Zang et al. 2021b). The gray "x" points in the right panels are the bad data automatically recognized by the updated algorithm ($\sigma_{\rm sub} > 2.5$ or $\sigma_{\rm res} > 2.0$). The preliminary data have more scatter and the errors are significantly underestimated. In the lower panels, the shaded regions $T_1$ and $T_4$ mark the most obvious anomalous regions in the TLC data (lower right). Similar anomalies also exist in the preliminary data (lower left) but are not significant compared to the systematic errors, e.g., $T_2$ and $T_3$, even if the noisy KMTA03 points are ignored.

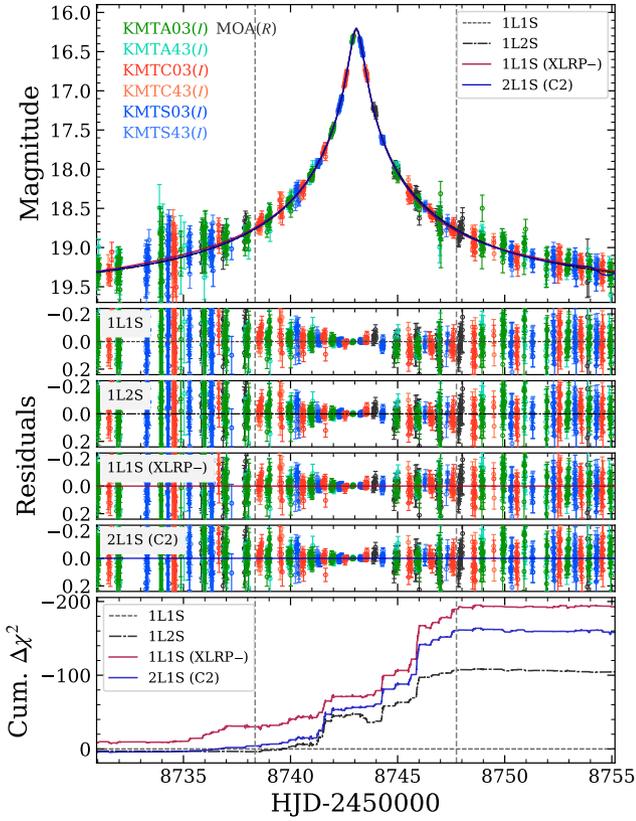

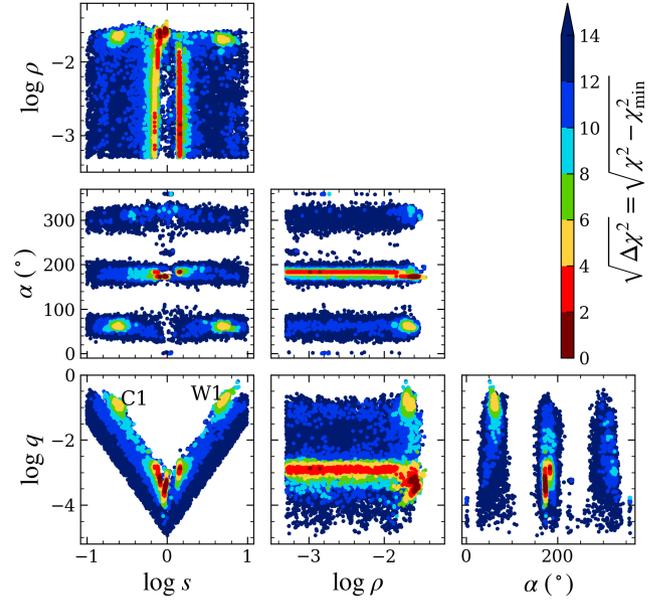

**Figure 7.** The hot MCMC results in the $(\log s, \log q, \alpha, \log \rho)$ space, the colors are coded by the $\Delta \chi^2$. The refined normal MCMCs over each local minima are overlapped on the hot chains. The two distinct solutions, C1 and W1 are marked. A zoom-in plot of the small-$q$ local minimums can be seen in Figure 8.

**Figure 5.** The light curve data of MOA-2019-BLG-421 around the peak together with the 1L1S, 1L2S, 1L1S + xallarap (XLRP), and 2L1S models. The residuals of each model are shown in separate panels. The lowest panel shows the cumulative $\Delta \chi^2$ relative to the standard 1L1S model.

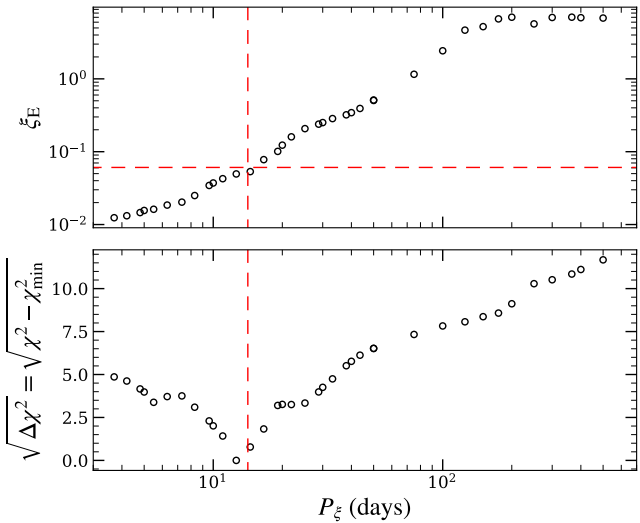

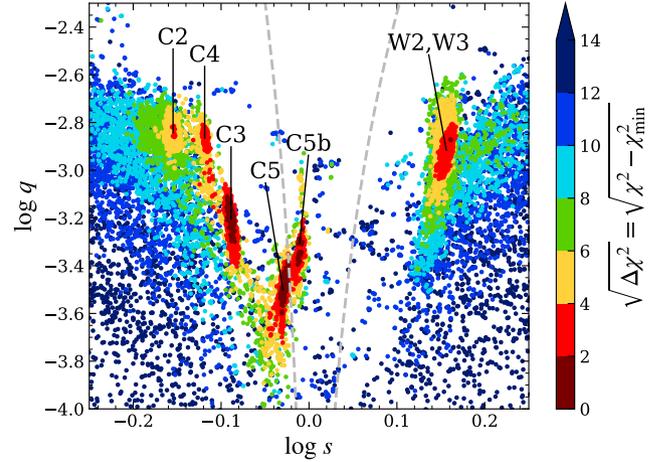

**Figure 8.** A zoom-in version of the lower left panel of Figure 7. All the small-$q$ solutions are marked. W2 and W3 occupy the same $(\log s, \log q)$ region but are different in the $\alpha$ space.

**Figure 6.** Results of the xallarap period $P_\xi$ search of the 1L1S model. Each point represents the optimized solution with the given xallarap period. The red lines indicate the values for the final best solution (XLRP−).

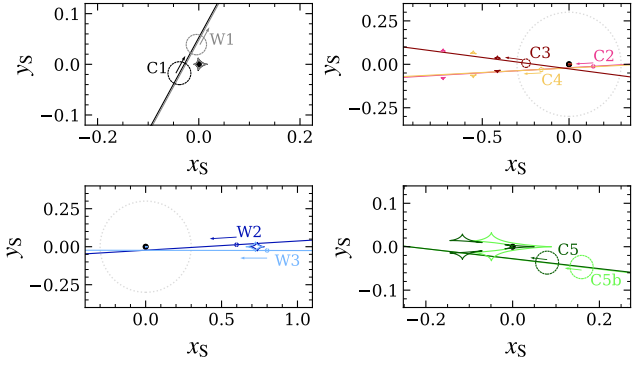

**Figure 9.** The topology of the caustic and the source trajectory on the lens plane for each 2L1S model.

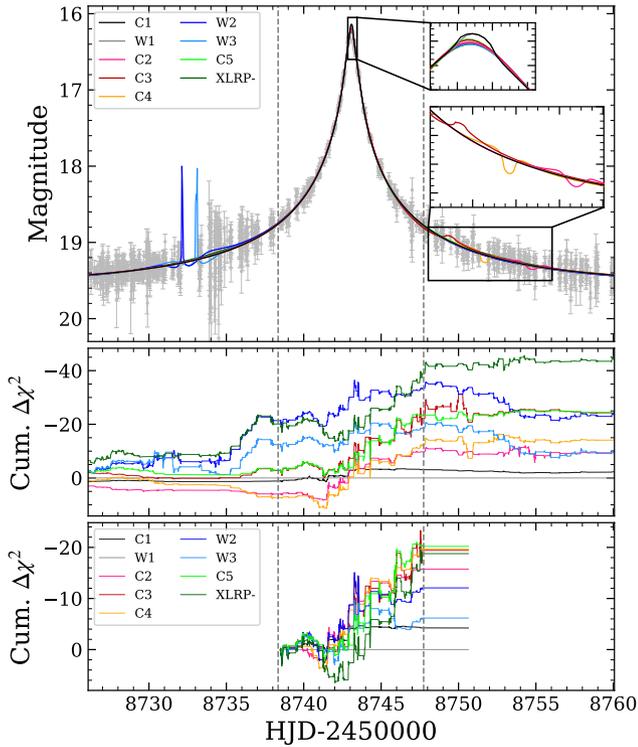

**Figure 10.** *(upper).* The light curve data of MOA-2019-BLG-421 around the peak together with all 2L1S models and the xallarap model. *(middle).* The cumulative $\Delta\chi^2$ when all data are included. *(lower).* The cumulative $\Delta\chi^2$ for only peak ($\sim t_0 \pm 4.5$ d) data.

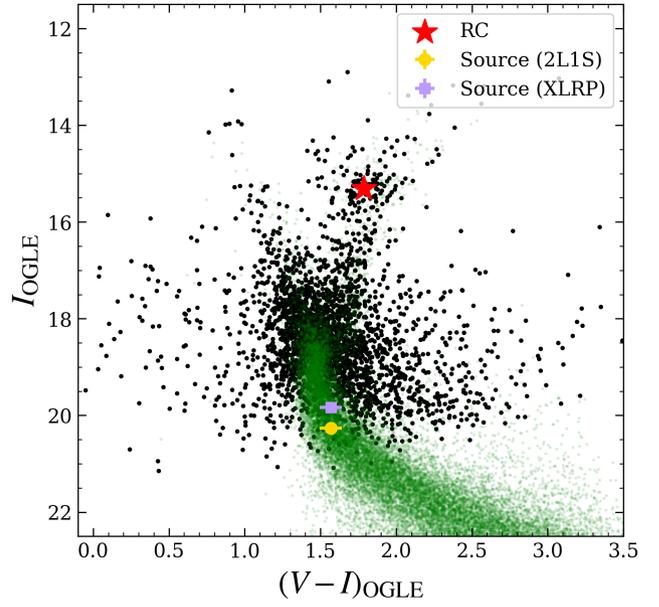

**Figure 11.** Color-magnitude diagrams (CMD) for the $2' \times 2'$ square field centered on MOA-2019-BLG-421. The black points are the field stars measured from KMTNet images, and they are calibrated to the OGLE-III color and magnitude (Szymański et al. 2011). Green points are from the CMD obtained by Holtzman et al. (1998) from HST observations of Baade's Window, which we have aligned to the KMTNet CMD using the centroid of the red clump. The positions of the red clump centroid (RC) and the microlens source for different interpretations are marked on the figure.

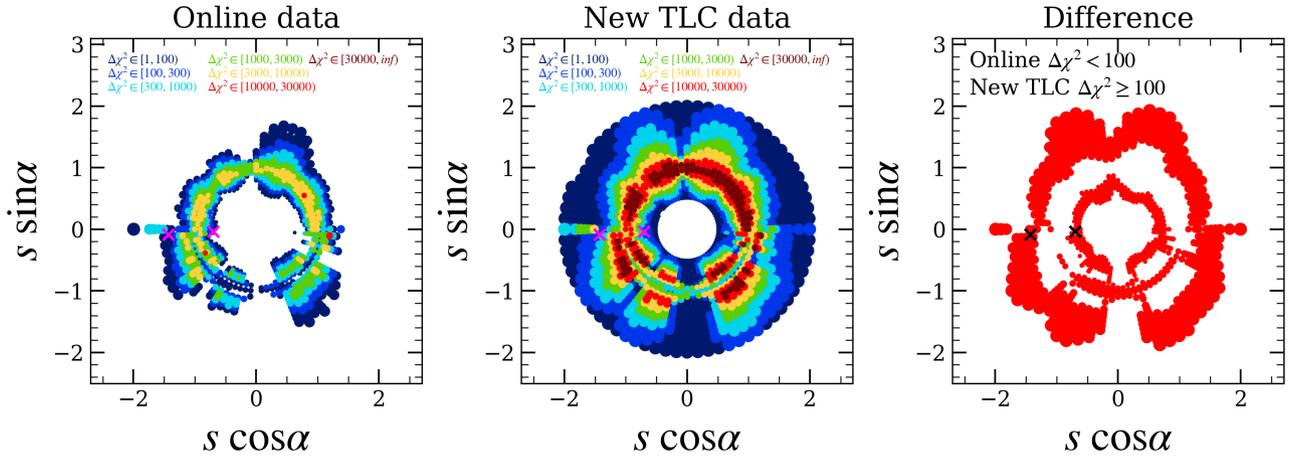

**Figure 12.** Planetary sensitivity of event MOA-2019-BLG-421 for a log $q = -2.8$ planet on the $(s, \alpha)$ plane. On the left two panels, each point represents an artificial light curve generated by the given $(s, q, \alpha)$ and the color represents the significance of the injected signal, i.e., $\Delta\chi^2 = \chi^2_{1L1S} - \chi^2_{1L1S}$. The rightmost panel shows the detection ($\Delta\chi^2 > 100$) rate enhanced by the new TLC data. The two crosses mark the two preferred 2L1S solutions (C2, W2).

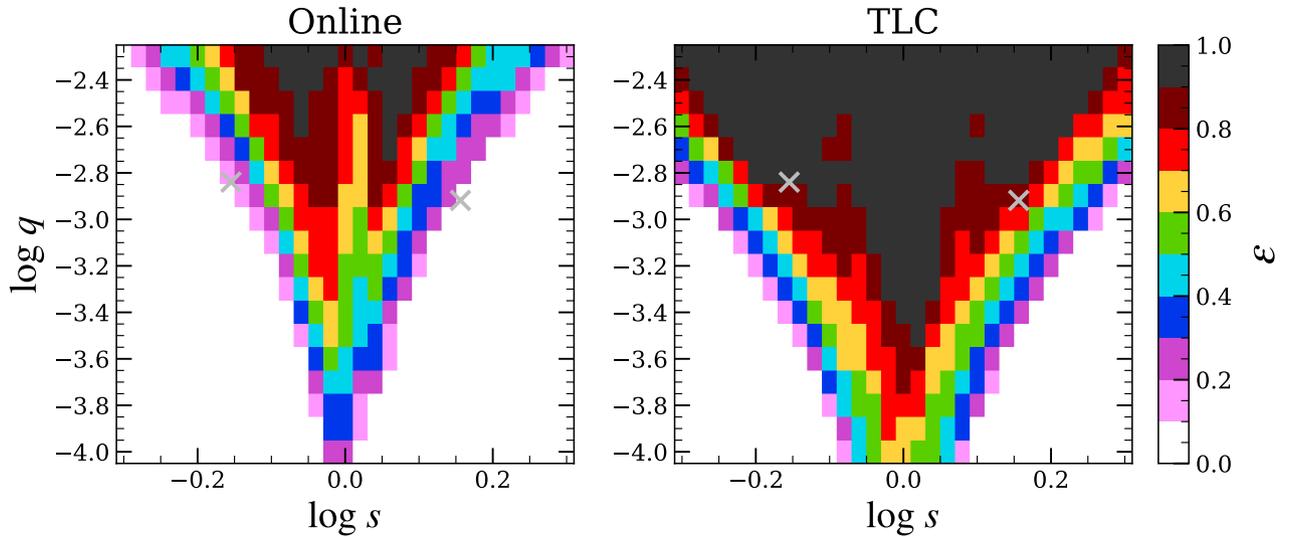

**Figure 13.** Planetary sensitivity for event MOA-2019-BLG-421 on the $(s, q)$ plane. The colors represent the fraction ($\varepsilon$) of a given $(s, q)$ planets that produce a deviation $\Delta\chi^2 > 100$. The gray crosses mark the two preferred 2L1S solutions (C2, W2).